\documentclass{jfm}
\usepackage{graphicx}
\usepackage{graphbox}
\usepackage{newtxtext}
\usepackage{newtxmath}
\usepackage{mathtools}

\usepackage{natbib}
\usepackage{hyperref}
\hypersetup{
    colorlinks = true,
    urlcolor   = blue,
    citecolor  = black,
}

\usepackage{multirow}
\usepackage{accents}
\usepackage{comment}
\usepackage{subcaption}
\usepackage{xcolor} 
\usepackage{epstopdf, epsfig}
\usepackage{tikz}

\newcommand{\RomanNumeralCaps}[1]
\linenumbers
\graphicspath{{figures/}}
\usepackage{cleveref}

\usepackage{graphicx} 
\usepackage{physics}
\usepackage{xcolor}
\usepackage{multirow}
\usepackage{cleveref}
\usepackage{array}
\usepackage{placeins}
\usepackage{subcaption}
\usepackage{amsmath,amsfonts,amssymb,stackengine,graphicx}
\usepackage{mathtools}

\newcommand{\T}{\vb{T}}

\newcommand{\C}{\vb{C}}
\newcommand{\I}{\vb{I}}
\newcommand{\U}{\vb{ U}}

\renewcommand{\u}{\vb{u'}}

\title{Revisiting 2D Viscoelastic Kolmogorov Flow: A Centre-mode-driven transition}

\author{Theo Lewy\aff{1}
  \corresp{\email{tal43@cam.ac.uk}}
 \and Rich Kerswell \aff{1}}

\affiliation{\aff{1} DAMTP, Centre for Mathematical Sciences, Wilberforce Road, Cambridge CB3 0WA, UK}

\begin{document}

\maketitle

\begin{abstract}

We revisit viscoelastic Kolmogorov flow to show that the elastic linear instability of an Oldroyd-B fluid at vanishing Reynolds numbers ($Re$) found by Boffetta et al. ({\em J. Fluid Mech.} {\bf 523}, 161-170, 2005) is the same `centre-mode' instability found at much higher $Re$ by Garg et al. ({\em Phys. Rev. Lett.} {\bf 121}, 024502, 2018) in a pipe and Khalid et al. ({\em J. Fluid Mech.} {\bf 915}, A43, 2021) in a channel. 
In contrast to these wall-bounded flows, the centre-mode instability exists even when the solvent viscosity vanishes (e.g. it exists in the upper-convective Maxwell limit with $Re=0$).
Floquet analysis reveals that the preferred centre-mode instability almost always has a wavelength twice that of the forcing. All elastic instabilities give rise to familiar `arrowheads’ (Page et al. {\em Phys. Rev. Lett.} {\bf 125}, 154501, 2020) which in sufficiently large domains and at sufficient Weissenberg number ($W$) interact chaotically in 2D to give elastic turbulence via a bursting scenario. Finally, it is found that the $k^{-4}$ scaling of the kinetic energy spectrum seen in this 2D elastic turbulence is already contained within the component arrowhead structures.

\end{abstract}

\section{Introduction}

Polymeric fluids such as plastic melts, oils, gels and paints are widespread across modern life. The presence of polymers introduces a myriad of phenomena that are not seen in simpler Newtonian fluids, due to the elasticity in the system. One key example of this is a chaotic self-sustaining state known as `elastic turbulence' (ET), which is driven by elastic effects and consists of dynamics across a range of length-scales \citep{Groisman2000}. The addition of just a small amount of polymer into a solvent can cause it to show signs of turbulence even when inertia is negligible, distinguishing it from Newtonian turbulence. 

ET was first identified in a curvilinear setting \citep{Groisman2000} in which hoop stresses were important in triggering the transition to turbulence \citep{Shaqfeh1996}. The later discovery of elastic turbulence in rectilinear geometries (e.g. experimentally by \cite{Pan2013} and \cite{Shnapp2022} in channel flow, \cite{Bonn2011} in pipe flow, and numerically by \cite{Berti2008} and \cite{Berti2010} in Kolmogorov flow, \cite{Rota2023} and \cite{Lellep2024} in channel flow, and \cite{Beneitez2023} in plane Couette flow) where linear hoop stress instabilities are absent, demonstrated that other mechanisms can also trigger ET. Two mechanisms have recently been uncovered.

The first is the `centre-mode' instability which has been identified in channel flow and pipe flow but is absent in plane Couette flow \citep{Garg_Chaudry_Khalid_Shankar_Subramanian_2018,Chaudhary_Garg_Subramanian_Shankar_2021,Khalid_Chaudhary_Garg_Shankar_Subramanian_2021, Khalid_Shankar_Subrmanian_2021}. Despite being entirely elastic in origin \citep{Buza2022a}, the instability only persists to vanishing Reynolds number in ultra-dilute channel flow \citep{Khalid_Shankar_Subrmanian_2021}. Significantly, the instability is subcritical in channel and pipe flow, causing instabilities at Weissenberg numbers $W$ much lower than those needed for linear instability \citep{Page2020, Dongdong2021,Buza2022a, Buza2022b, Morozov2022}. Solutions on the upper branch resemble `arrowheads' \citep{Page2020}, and these arrowhead structures can be identified within elasto-inertial turbulence in 2D channel flow when inertia is not neglected \citep{Dubief2022,Beneitez_Page_Dubief_Kerswell_2024}. 

The second new mechanism is another linear instability called the `polymer diffusive instability' (PDI) which has been found very recently in plane Couette, channel and pipe flow localised at the boundaries \citep{Beneitez2023, Couchman2024, Lewy2024}. It exists in inertialess systems, and requires the presence of polymer stress diffusion, either explicitly included in the model or  implicitly applied  by the time-stepping numerical scheme used. It has been shown to lead to a chaotic state in 3D channel flow, hinting at its relevance in the transition to wall bounded elastic turbulence \citep{Beneitez2023, Beneitez2024b, Rota2023}. 

In light of these developments, it seems worthwhile to revisit viscoelastic Kolmogorov flow (vKf) which was studied  \citep{BOFFETTA_CELANI_MAZZINO_PULIAFITO_VERGASSOLA_2005} over a decade before the centre-mode instability was announced by \cite{Garg_Chaudry_Khalid_Shankar_Subramanian_2018} in pipe flow (at very different $Re$) and nearly 2 decades before PDI was found \citep{Beneitez2023}. The original linear analysis by \cite{BOFFETTA_CELANI_MAZZINO_PULIAFITO_VERGASSOLA_2005} took the form of a multiscale analysis conducted at low $Re$ ($\lesssim 6$) over multiple forcing wavelengths but did not plot any unstable eigenmodes. Later numerical simulation work at  $Re \lesssim 1$ by \cite{Berti2008} and \cite{Berti2010}, however, clearly shows  arrowhead structures indicative of the centre-mode instability. A recent asymptotic analysis of the centre-mode instability \cite{Kerswell_2023} confirms  that it exists in inertialess vKf 
but doesn't exclude the presence of other instabilities. In particular, the key ingredient for PDI may actually be maxima of the base shear, which vKf has, rather than boundaries per se, which it doesn't \citep{Lewy2024}. Our objectives are therefore to: i) carry out a full investigation of the linear stability problem including over multiple forcing wavelengths using Floquet analysis; ii) clarify whether PDI exists or indeed any other elastic or elasto-inertial instability occurs beyond the known Newtonian instability; and iii) then, armed with this information, explore the transition scenarios possible.


The structure of this paper is as follows. In \cref{formulation} the governing equations to be used are introduced, the 1D base state identified and the associated linearised equations derived. The results of solving the linear instability eigenvalue problem are then discussed in \cref{linear_instability}, where the centre-mode instability can be identified and  PDI is shown to be absent. The centre-mode is found to exists across a wide range of parameters in this system including an inertialess upper-convected Maxwell fluid. We identify perturbations with a period that is twice the forcing periodicity to be the linearly most unstable using Floquet analysis. In addition, we show that the flow relaminarises at sufficiently large $W$ for a fixed geometry. We then move on to non-linear behaviour, identifying the centre-mode as subcritical in \cref{subcriticality}, as well as plotting a range of stable exact coherent structures. Section \ref{turbulence} considers ET, suggesting that a bifurcation of the centre-mode instability is the cause of this chaos, and shows that the onset of turbulence is marked by the presence of bursting solutions. In addition, we see that the power spectra and energy budget of a simple periodic arrowhead solution is similar to that of ET. A final discussion is given in \cref{conclusion}.


\section{Formulation} \label{formulation}

\begin{figure}
\centering
\includegraphics[width=0.6\textwidth]{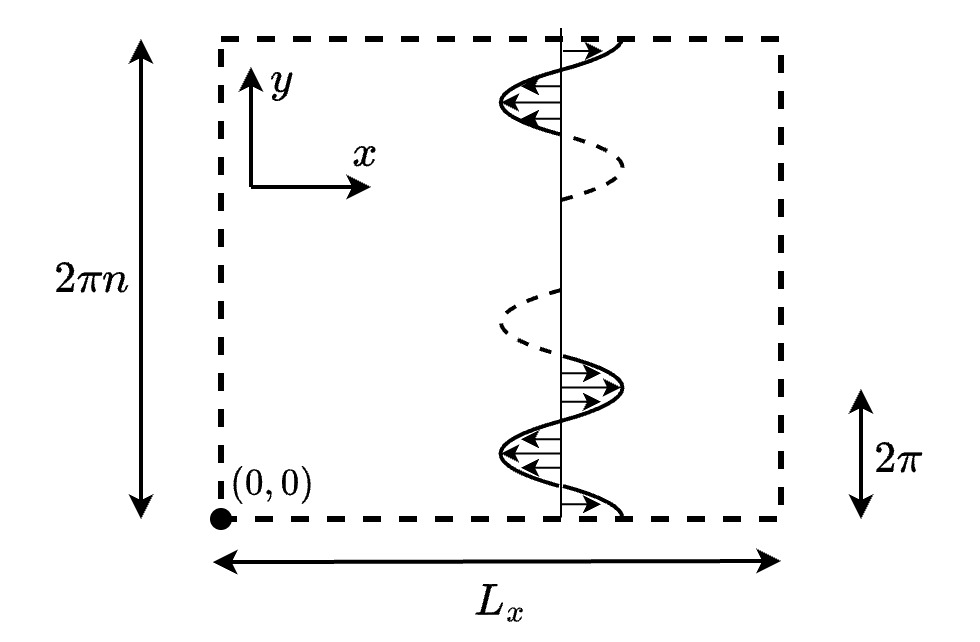}
\caption{The Kolmogorov flow setup with forcing wavelength $2\pi$. Perturbations have wavelength $2\pi n$ in the $\vb {\hat y}$ direction (where $n$ is an integer) and $L_x$ in the $\vb {\hat x}$ direction.}  \label{Kolmogorov_diagram}
\end{figure}

We consider 2D Kolmogorov flow of an Oldroyd-B fluid with forcing in the $\vb {\hat x}$ direction that varies periodically with $\vb {\hat y}$ in the perpendicular direction. The non-dimensionalised equations relating the velocity $\vb u$, the polymeric stress tensor $\vb T$, the conformation tensor $\vb C$ and the pressure $p$ are
\begin{equation}
\begin{gathered}
    Re (\frac{\partial \vb u}{\partial t} + \vb u \cdot \nabla \vb u) = - \nabla p +  (1-\beta) \nabla \cdot \vb{T} + \beta \nabla^2 \vb{u}  + \left(\frac{1+\varepsilon \beta W}{1+\varepsilon W}\right) \cos y \, \vb{\hat x},
\end{gathered}\label{governing1}
\end{equation}
\begin{equation}
\begin{gathered}
    \frac{\partial{\C}}{\partial t} + \vb u \cdot \nabla \C - \nabla \vb u^T  \cdot \C - \C \cdot \nabla \vb u + \T = \varepsilon \nabla ^2 \C,
\end{gathered}\label{governing2}
\end{equation}
\begin{equation}
\begin{gathered}
    \nabla \cdot \vb u = 0,
\end{gathered}\label{governing3}
\end{equation}
and
\begin{equation}
\begin{gathered}\label{governing4}
\T = \frac{1}{W}(\C-\I).
\end{gathered}
\end{equation}
We consider a domain of size $[0, L_x] \cross [0, 2\pi n]$, where $L_x$ is the horizontal extent, and the integer $n$ is the number of forcing wavelengths applied to the system: see \cref{Kolmogorov_diagram}. Periodic boundary conditions are imposed in both directions so in particular, flows with a wavelength $n$ times longer than the forcing are permitted. All variables are scaled using the laminar peak velocity $U_0$, the total kinematic viscosity $\nu = \nu_s + \nu_p$, which is the sum of the solvent and polymer viscosities respectively, and the lengthscale $L_0/2\pi$, where $L_0$ is the forcing wavelength. The coefficient of the forcing term in \cref{governing1} is to ensure that the resulting base velocity has unit amplitude.

%
%

%
%
These equations use dimensionless parameters
\begin{equation} 
Re \coloneqq \frac{U_0 L_0}{2 \pi \nu}, \qquad W \coloneqq \frac{2 \pi U_0 \lambda  }{L_0} \label{Re&W}
\end{equation}
\begin{equation}
\beta \coloneqq \frac{\nu_s}{\nu}, \qquad \varepsilon \coloneqq \frac{2 \pi \delta}{U_0 L_0}
\label{beta}
\end{equation}
where $\lambda$ is the relaxation time and $\delta$ is the polymer stress diffusion coefficient. It will be useful to define the Elasticity number 
\begin{equation}
E \coloneqq \frac{W}{Re}
\end{equation}
which is the measure of elasticity used to introduce the centre-mode instability \citep{Garg_Chaudry_Khalid_Shankar_Subramanian_2018, Khalid_Chaudhary_Garg_Shankar_Subramanian_2021, Chaudhary_Garg_Subramanian_Shankar_2021}. The Oldroyd-B model reduces to the upper-convected Maxwell (UCM) when the concentration $\beta=0$. 

\subsection{Symmetries}\label{symmetries}

The above system has 3 types of symmetries associated with it, like in the Newtonian case \citep{Chandler_Kerswell_2013}. The shift-reflect symmetry maps
\begin{equation}
\mathcal{S} [u, v, T_{xx}, T_{xy}, T_{yy}, p](x, y) \rightarrow [-u, v, T_{xx}, -T_{xy}, T_{yy}, p](-x, y+\pi)
\end{equation}
where $\vb u = u \vb{\hat x} + v \vb{\hat y}$ and $\vb T = T_{xx} \vb{\hat x}\vb{\hat x} + T_{xy} (\vb{\hat x}\vb{\hat y} + \vb{\hat y}\vb{\hat x}) + T_{yy}\vb{\hat y}\vb{\hat y}$. A reflection symmetry maps 
\begin{equation}
\mathcal{R}[u, v, T_{xx}, T_{xy}, T_{yy}, p](x, y) \rightarrow [u, -v, T_{xx}, -T_{xy}, T_{yy}, p](x, 2\pi-y).
\end{equation}
In addition to these 2 discrete symmetries, there is the continuous translational symmetry
\begin{equation}
\mathcal{T}_s[u, v, T_{xx}, T_{xy}, T_{yy}, p](x, y) \rightarrow [u, v, T_{xx}, T_{xy}, T_{yy}, p](x+s, y)
\end{equation}
for $s \in [0, L_x)$. Every solution is therefore associated with a set of solutions generated via these symmetries, and in particular the shift-reflect symmetry means that any solution moving in the positive $\vb{\hat x}$ direction has an associated solution moving in the negative $\vb{\hat x}$ direction. 



\subsection{Base flow}

There is a one-dimensional base flow which depends only on $y$ which is 
$$ \vb U = \cos y \, \vb{\hat x}, \quad P=0, $$
$$ T_{xx} = \frac{W}{1+\varepsilon W} \left(1 - \frac{\cos 2y}{1+4\varepsilon W}\right), \quad T_{xy} = \frac{-1}{1+\varepsilon W} \sin y \quad \& \quad T_{yy} = 0.$$

\subsection{Linearising}

To examine the linear stability of the base state, small perturbations proportional to $e^{ik(x-ct)}$ of all dependent variables are considered where $k\in \mathbb{R}$ is the wavenumber, and $c \in \mathbb{C}$ is an eigenvalue to be found. This results in variables of the form
\begin{align}\label{linearising_ansatz}
 \vb{u} &=  \vb U(y)+ \begin{bmatrix} u'(y) \\ v'(y) \end{bmatrix}
 e^{ik(x-ct)}, 
 \quad
  p =  p'(y)e^{ik(x-ct)} 
 \nonumber \\ 
 \vb{T} &= \vb T(y) + \begin{bmatrix} \tau_{xx}'(y) & \tau_{xy}'(y) \\ \tau_{xy}'(y) & \tau_{yy}'(y) \end{bmatrix}e^{ik(x-ct)}
\end{align}
where a dash denotes a perturbative quantity that, due to the boundary conditions, must be $2\pi n$ periodic in $y$. The imaginary part of the eigenvalue, $c_i$, determines the linear stability of the system, with $\sigma \coloneqq kc_i$ the growth rate. To reduce slightly the linearised equations which determine the evolution of the perturbations, we take the curl of (\ref{governing1}) to eliminate the pressure, and use (\ref{governing4}) to write all $\C$ in terms of $\T$. Equations (\ref{governing1})-(\ref{governing4}) then become

\begin{align}\label{linearised1}
 ikRe\left[ (U-c)(D^2-k^2) - D^2U\right]v' = &-(1-\beta)\left[ -k^2 D(\tau_{xx}' - {\tau_{yy}'}) +  ik(D^2 + {k^2}) \tau_{xy}' \right] \nonumber\\ &+ \beta( D^2 - {k^2})^2 v',
\end{align}
\begin{align}\label{linearised2}
\begin{gathered}
\left[ik(U-c) + \frac{1}{W}\right]\tau_{xx}' = -v'DT_{xx} + 2ikT_{xx} u' + 2T_{xy} Du' + 2\tau_{xy}'DU + {\frac{2ik}{W}u'}
\end{gathered},
\end{align}
\begin{align}\label{linearised3}
\begin{gathered}
\left[ik(U-c) + \frac{1}{W}\right]\tau_{xy}' = -v'DT_{xy} + ikT_{xx} v' + \tau_{yy}'DU + \frac{1}{W}(Du' + {ikv'})
\end{gathered},
\end{align}
\begin{align}\label{linearised4}
\begin{gathered}
\left[ik(U-c) + \frac{1}{W}\right]\tau_{yy}' = 2ikT_{xy} v' + \frac{2}{W}Dv'
\end{gathered},
\end{align}
\begin{align}\label{linearised5}
\begin{gathered}
iku' + Dv' = 0
\end{gathered}.
\end{align}
where $D\coloneqq d/dy$. The costly procedure of solving this eigenvalue problem over the whole domain  $y\in[0,2\pi n]$ can be avoided by applying Floquet analysis just across one forcing wavelength $y\in[0,2\pi]$ and including a modulation parameter $\mu$ to compensate. A mode with Floquet exponent $\mu$ has perturbations of the form $\phi' = \hat \phi(y) e^{i\mu y}$ where $\phi'$ is the perturbation of any flow variable, and $\hat \phi$ is $2\pi$ periodic. The resultant perturbation has periodicity $2\pi / \mu$ when $1/\mu \in \mathbb{N}$, as all base flow quantities have the same periodicity as the forcing. Values of $1/\mu$ which factor into $n$ then satisfy the periodic boundary conditions over the large domain of  $y\in[0,2\pi n]$.

\section{Linear Instability} \label{linear_instability}

In this section, the linear instability seen in viscoelastic Kolmogorov flow by \citet{BOFFETTA_CELANI_MAZZINO_PULIAFITO_VERGASSOLA_2005} is identified as the centre-mode instability \citep{Garg_Chaudry_Khalid_Shankar_Subramanian_2018, Chaudhary_Garg_Subramanian_Shankar_2021, Khalid_Chaudhary_Garg_Shankar_Subramanian_2021}. We begin by considering vKf when the flow has the same spatial periodicity as the forcing  (i.e. $n=1$), and see that: i) the instability scales with $E$ like the centre-mode in a channel when $E\ll1$; and ii) the eigenfunction resembles the centre mode. We then consider $E\gg1$, as well as how increasing $n$, the number of forcing wavelengths, affects the instability. The centre-mode instability is not confined to dilute vKf, but is found across all $\beta \in [0,1)$, even existing for a UCM fluid with $\beta=0$. Curiously, the flow is also found to restabilise as $W\rightarrow \infty$ within a geometry of fixed streamwise extent. All numerics were computed using the open-source software Dedalus \citep{Burns2020}.

\subsection{Harmonic Disturbances ($n=1$)}\label{n=1}

\citet{Kerswell_2023} show that there is an unstable eigenfunction when $n=1$ in ultra-dilute vKf that resembles the centre-mode eigenfunction in a channel. Here we go a step further and show that the $(Re,k)$ neutral curves show the distinctive centre-mode loops seen in channel flow (as shown in fig. 10 of \cite{Khalid_Chaudhary_Garg_Shankar_Subramanian_2021}), and they follow the same scaling relation for $E \ll 1$. However, we also show that the behaviour for large $E$ is substantially different in this unbounded flow to the centre-mode in channel flow, as the instability is not suppressed as $E\rightarrow \infty$. Instead, the instability exists in inertialess vKf.

\begin{figure}
\centering
\includegraphics[width=0.65\textwidth]{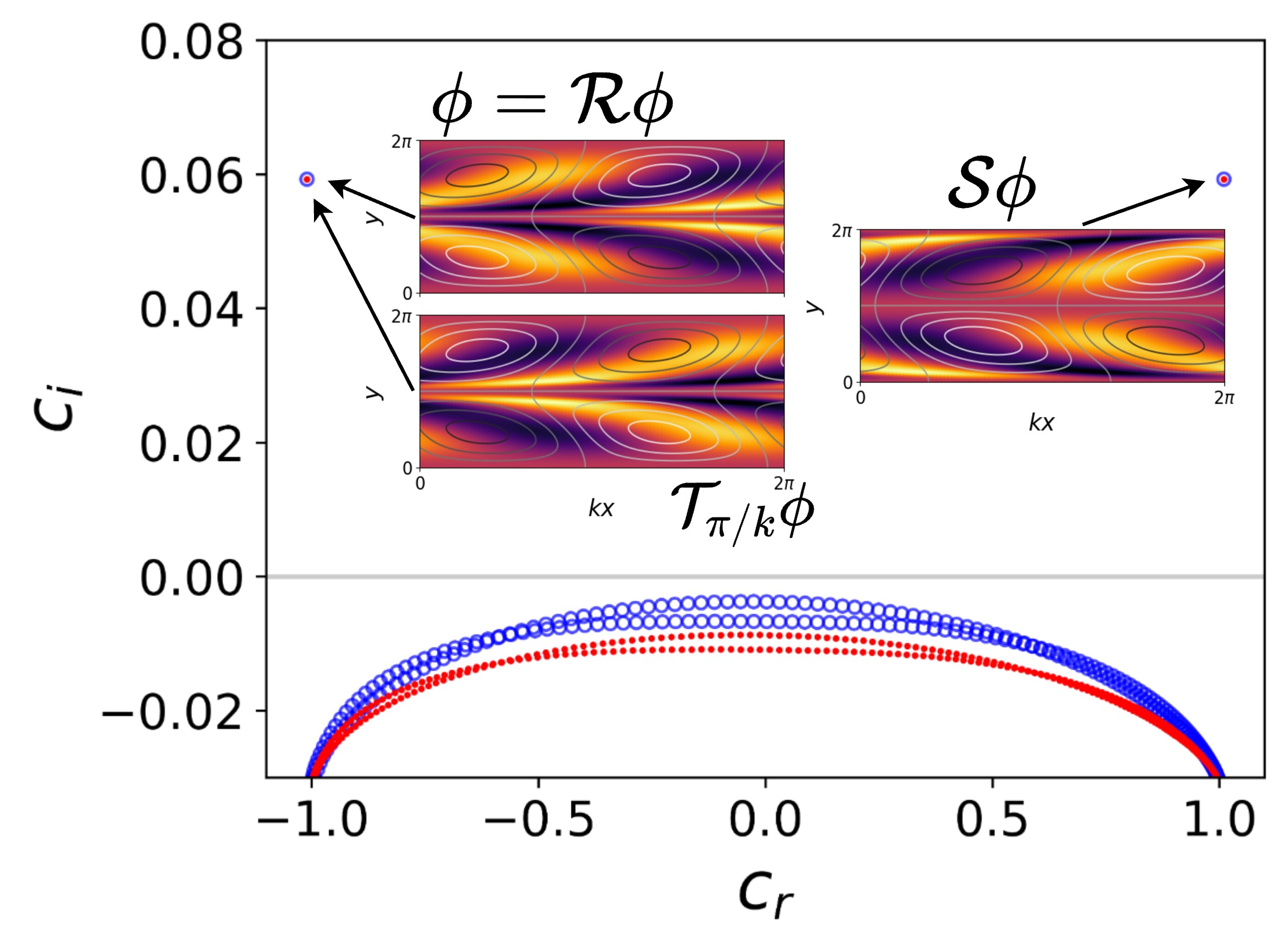}
\caption{The eigenvalue spectrum when $E=81$, $Re=2$, $\beta=0.95$, $\varepsilon=0$, $\mu=0$ and $k=0.2$, with resolution $N_y=300$ (blue circles) and $N_y=400$ (red dots). The centre-mode has unstable eigenvalues at $c=\pm1.01016736+0.05925964j$, while a stable continuous spectra is seen with $c_i<0$. Insets show the polymer stress trace (colours) and streamfunction (contours) of an eigenmode $\phi$, alongside symmetries of $\phi$. While reflections $\mathcal{R}$ and translations $\mathcal{T}_s$ leave the eigenvalue $c=c_r+ic_i$ unchanged, shift-reflections $\mathcal{S}$ produce modes with eigenvalue $-c_r+ic_i$ that travel in the opposite direction. }\label{evalue_spectrum}
\end{figure}

To consider the system with $n=1$ we take the linearised system with Floquet exponent $\mu = 0$. We plot an example eigenvalue spectra in \cref{evalue_spectrum}, showing how an eigenmode and its eigenvalue $c=c_r+ic_i$ are affected by the symmetries discussed in \cref{symmetries}. The shift-reflect symmetry $\mathcal{S}$ generates a new eigenmode with eigenvalue $c=-c_r+ic_i$ that travels in the opposite direction. The symmetries of the centre-mode in the $n=1$ case make the eigenmode invariant under reflection $\mathcal{R}$. Lastly, translational symmetries $\mathcal{T}_s$ correspond to a phase shift of the mode. The stability of all such symmetries is the same, as the growth rate is unchanged under all operators.

\begin{figure}
\centering
\includegraphics[width=0.85\textwidth]{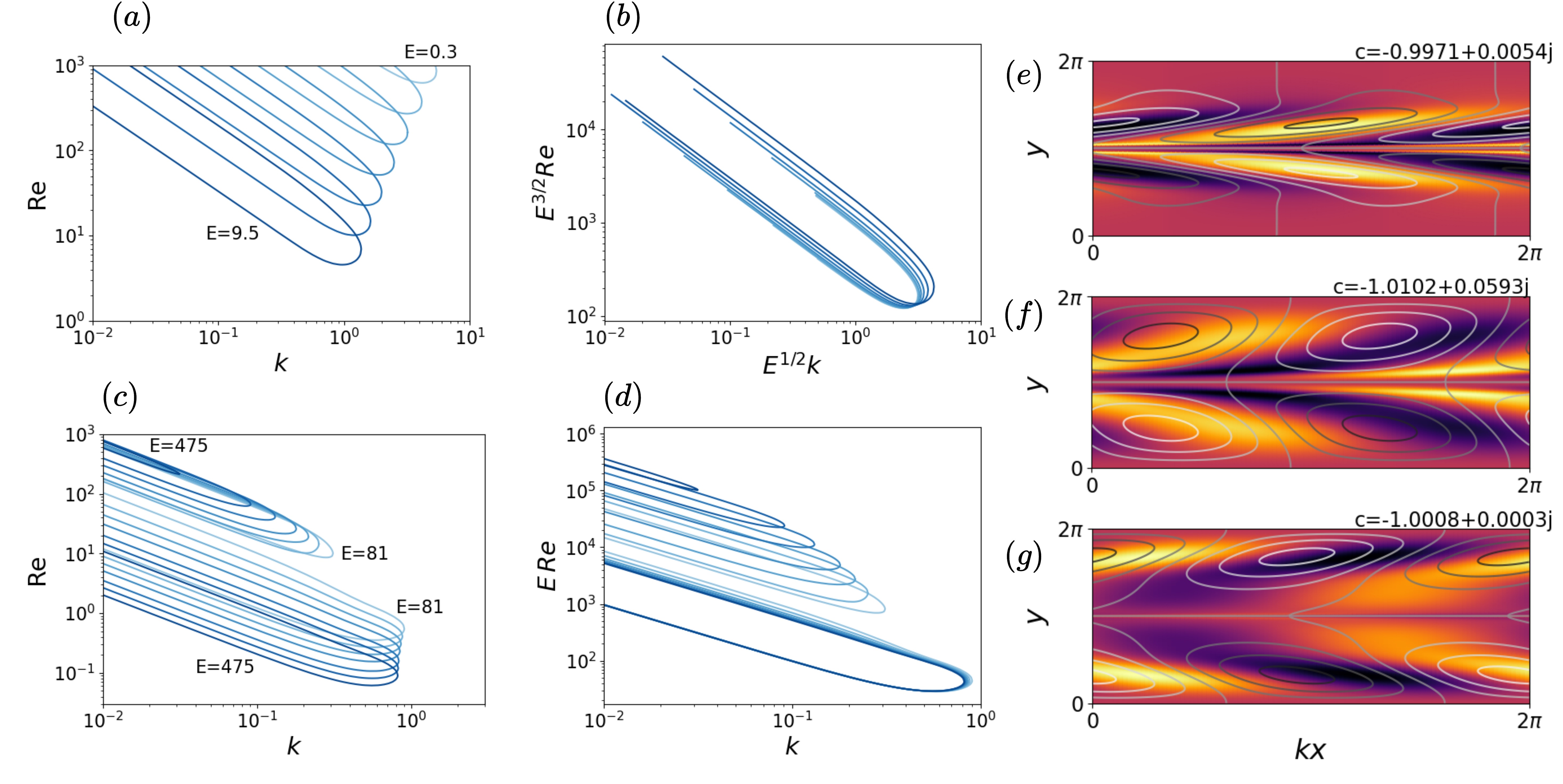}
\caption{$a-d)$ The centre-mode neutral curves in the $(Re, k)$ plane for $\beta=0.95$, $\varepsilon=0$, $\mu=0$ and $a, b)$ $E = 0.3, 0.6, 1.0, 1.8, 3.1, 5.5, 9.5$ (light to dark) $c, d)$ $E=81, 107, 142, 187, 272, 359, 475$. (light to dark). Note $b)$ and $d)$ are scaled versions of $a)$ and $c)$ respectively, demonstrating that for small $E$, $Re_{crit} \sim E^{-3/2}$, while for large $E$, $Re_{crit} \sim E^{-1}$. We plot eigenfunctions with $k=0.2$ and $e)$ $E=3.1, Re=300$, $f)$ $E=81, Re=2$ and $g)$ $E=81, Re=20$. These correspond to the instability in the low $E$ regime, the main loop in the high $E$ regime, and the secondary loop in the high $E$ regime respectively. Colours show the polymer stress trace field, while contours show the streamfunction. This figure demonstrates that the elastic instability seen at high $\beta$ and low $E$ is the centre-mode, and that a different scaling regime exists at high $E$.}\label{Re-k_neutral_curves}
\end{figure}

Figures \ref{Re-k_neutral_curves}a and \ref{Re-k_neutral_curves}b show unscaled and scaled $(Re,k)$ neutral curves for small values of $E$. The neutral curves takes the form of loops at small $E$, as is the case in channel flow. For these loops, $Re_{crit} \sim E^{-3/2}$ and $k_{crit} \sim E^{-1/2}$ where $Re_{crit}$ denotes the smallest Reynolds number on the neutral curve and $k_{crit}$ denotes the wavenumber at that point. These match the scalings for the centre-mode in channel flow \citep{Khalid_Chaudhary_Garg_Shankar_Subramanian_2021}. An eigenfunction in this regime is plotted in \cref{Re-k_neutral_curves}e, resembling that of the channel flow centre-mode as seen in fig. 5 of \citet{Khalid_Chaudhary_Garg_Shankar_Subramanian_2021}. The scaling relation and the eigenfunction both suggest that this instability is the centre-mode.


Figures \ref{Re-k_neutral_curves}c and \ref{Re-k_neutral_curves}d show the neutral curves for large $E$. The behaviour here is substantially different to that of channel and pipe flow, where at sufficiently large $E$ the centre-mode is suppressed \citep{Khalid_Chaudhary_Garg_Shankar_Subramanian_2021, Chaudhary_Garg_Subramanian_Shankar_2021}. Here in vKf, the instability exists at all $E$ (only up to $E=475$ is shown), and we see that the centre-mode loops have $Re_{crit} \sim E^{-1}$ and $k_{crit} \sim E^0$ as $E \rightarrow \infty$. Equivalently one can consider the scalings in terms of $Re$ and $W$ to obtain that as $Re \rightarrow 0$ for fixed $W$, $W_{crit} \sim Re^0$ and $k_{crit} \sim Re^0$, meaning these critical parameters are independent of $Re$. While $Re_{crit}$ and $k_{crit}$ are on a loop with these scalings, a secondary loop also exists at larger $Re$. These secondary loops collapse as $E$ increases. It is worth pointing out that in this regime, the largest streamwise wavenumber at which instability is seen is $k \approx 0.9$, and hence no linear instability is seen when simulating a box with $L_x=2\pi$ and $n=1$, as is true in the Newtonian case \citep{Marchioro1986}. The eigenfunctions for large $E$ are seen in figures \ref{Re-k_neutral_curves}f and \ref{Re-k_neutral_curves}g, showing the main loop and the secondary loop respectively. The eigenfunctions on the secondary loop visually resemble that of the main loop, though activity is more spread out across the flow and less clearly localised to $y=\pi$.

%
%

\begin{figure}
\centering
\includegraphics[width=0.85\textwidth]{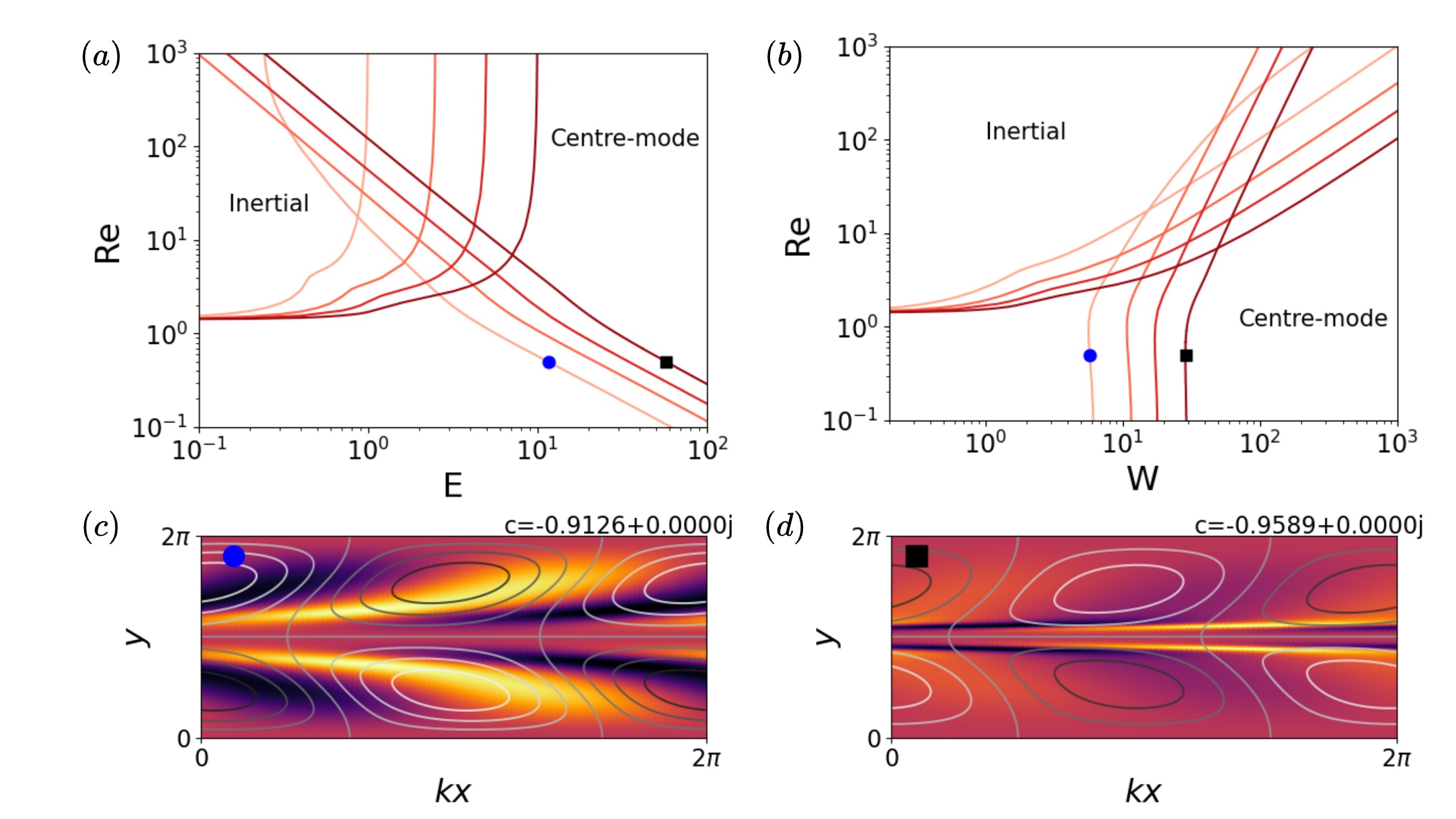}
\caption{The neutral curves across $k\in \mathbb{R}$ in the $a)$ $(Re, E)$ and $b)$ $(Re, W)$ plane when $\varepsilon=0$, $\mu=0$ and $\beta=0.5, 0.8, 0.9, 0.95$ (light to dark). This demonstrates that the centre-mode exists in the inertialess system across a range of $\beta$. Eigenfunctions for parameters on the neutral curves are shown in $c)$ when $(\beta, Re, W, kx)=(0.5, 0.5, 5.78, 0.47)$ (blue circle) and $d)$ when $(\beta, Re, W, kx)=(0.95, 0.5, 28.7, 0.60)$ (black square). Colours show the polymer stress trace field, while contours show the streamfunction.}  \label{Re-WE_neutral_curves}
\end{figure}

The neutral curves in the $(Re, E)$ and $(Re, W)$ plane are shown in \cref{Re-WE_neutral_curves}. These demonstrate that the centre-mode instability is linearly unstable at vanishing $Re$ number in vKf over a range of concentrations $\beta$. This contrasts with the cases of inertialess pipe flow, which is linearly stable \citep{Chaudhary_Garg_Subramanian_Shankar_2021} and inertialess channel flow, which is only linearly unstable for ultra-dilute fluids with $\beta > 0.9905$ \citep{Khalid_Chaudhary_Garg_Shankar_Subramanian_2021, Khalid_Shankar_Subrmanian_2021}. 

The neutral curves also show a second instability, which exists at vanishing $W$ and is inertial in nature. This instability is seen in Newtonian Kolmogorov flow, with a purely imaginary eigenvalue $c$ \citep{Meshalkin1961} and we identify an instability as inertial here if it similarly has zero frequency. This instability is suppressed by elasticity, with there being a maximum $E$ at which it exists. 

%
%

%
%
\subsection{The absence of the polymer diffusive instability}\label{PDI}

\begin{figure}
\centering
\includegraphics[width=0.85\textwidth]{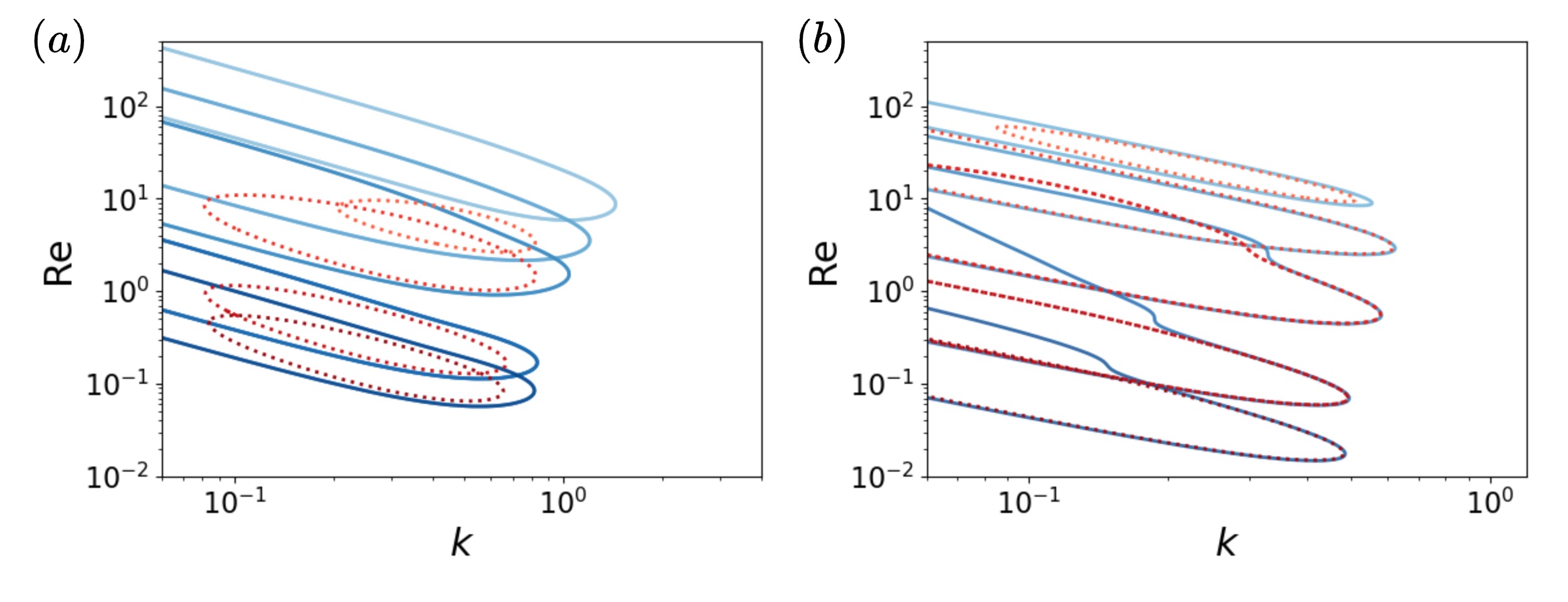}
\caption{The neutral curves for non-zero-frequency modes in the $(Re, k)$ plane when $\mu=0$, $\varepsilon=0$ (blue solid lines) and finite $\varepsilon=10^{-3}$ (red dotted lines) when a) $\beta=0.95$ and $E = 8, 16, 32, 256, 512$ (light to dark) and b) $\beta=0.2$ and $E=0.5, 1, 2, 8, 64, 256$. (light to dark). Wavenumbers as high as $k=100$ were considered. These demonstrate that the polymer diffusive instability was not identified in Kolmogorov flow, and that finite $\varepsilon$ generally stabilises the centre-mode instability.}  \label{finite_eps_neutral_curves_Re_kx}
\end{figure}

The linear stability analysis in section \ref{n=1} was performed with$\varepsilon=0$, and hence PDI is absent. However, to run simulations, introducing a finite $\varepsilon$ could potentially introduce PDI, as is the case for wall-bounded flows. We check that PDI is not in this system by considering how the neutral curves in the $(W,k)$ plane are affected by introducing finite $\varepsilon=10^{-3}$ in \cref{finite_eps_neutral_curves_Re_kx}. Wavenumbers up to $k=100$ were considered. The centre-mode loops that exists with vanishing $\varepsilon$ are adjusted slightly, but no new modes of instability are identified, meaning PDI was not found in vKf. This was true when $\beta=0.95$, meaning the simulations in sections \ref{subcriticality} and \ref{turbulence} do not contain PDI. In addition, we checked for PDI in a more concentrated fluid ($\beta=0.2$), where PDI was demonstrated to be particularly unstable in bounded flows \citep{Lewy2024}. The lack of PDI in Kolmogorov flow is consistent with \citet{Lewy2024} which considers $\beta \ll 1$ and suggests that PDI requires boundaries to exist in such a fluid. This finding confirms that  PDI is not necessarily seeded at regions of maximal base shear when polymer stress diffusion is present, as was the case in the wall-bounded rectilinear flows.

%
%

%
%
\subsection{Modulated Disturbances ($n>1$)}\label{n>1}

\begin{figure} 
\centering
\includegraphics[width=0.85\textwidth]{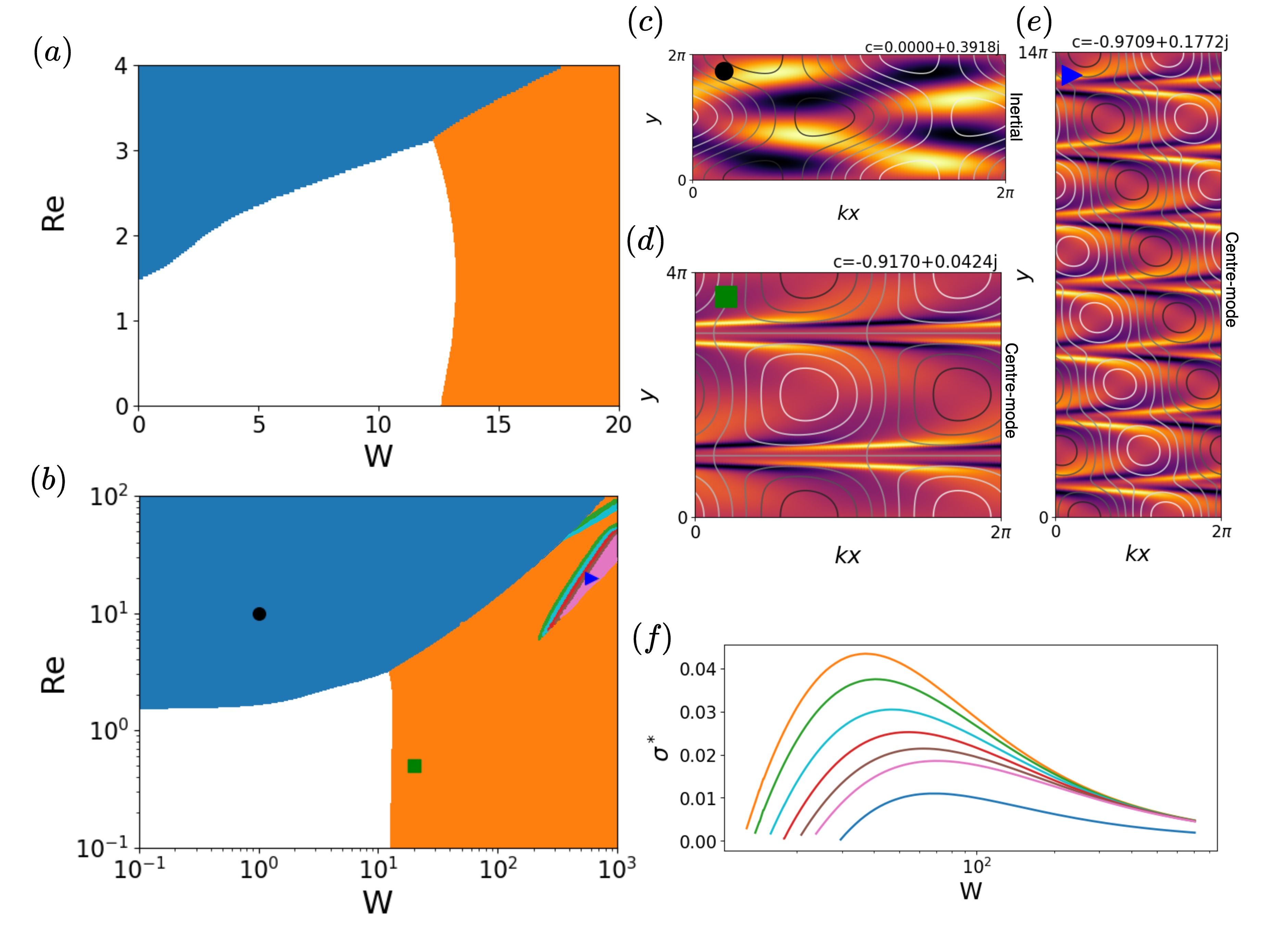}
\caption{$(a)$ The most unstable Floquet modes in the $(Re, W)$ plane for $\beta=0.95$, with $\varepsilon=0$ and instabilities over wavenumbers $k \in \mathbb{R}$ are considered, with colour denoting which Floquet mode is most unstable. Colours correspond to $\mu=0$ (blue), $\mu=1/2$ (orange), $\mu=1/3$ (green), $\mu=1/4$ (cyan), $\mu=1/5$ (red), $\mu=1/6$ (brown), $\mu=1/7$ (pink). $(b)$ The same on a log scale. Eigenfunctions are plotted with parameters $(c)$ $(W, Re, k, \mu) = (1, 10, 0.5, 0)$, $(d)$ $(W, Re, k, \mu) = (20, 0.5, 0.5, 1/2)$ and $(e)$ $(W, Re, k, \mu) = (600, 20, 0.01, 1/7)$. $(f)$ The maximum growth rate $\sigma^*$ of each Floquet mode (same colours as in $a)$) across all $k\in \mathbb{R}$ for $Re=0$, $\beta=0.95$ and $\varepsilon=0$ as $W$ varies. These plots demonstrate that all elastic instabilities are the centre-mode, which is generally most unstable when $\mu=1/2$, while the inertial instability is most unstable when $\mu=0$.}  \label{floquet}
\end{figure}

In this section we consider linear stability when $n>1$ by considering modes with Floquet exponents $\mu$ satisfying $1/\mu \in \mathbb{N}$. 
Figure \ref{floquet} shows the most unstable Floquet modes in the $(Re, W)$ plane for a viscosity ratio of $\beta=0.95$. All wavenumbers $k\in \mathbb{R}$ are considered, as well as the first 7 Floquet modes in figures \ref{floquet}a and b, which focus on the behaviour of the system at small and large $(Re, W)$ respectively. The orange regions demonstrate that it is the $\mu=1/2$ Floquet mode that generally makes the centre-mode instability most unstable, corresponding to perturbations with wavelengths of $4\pi$ in the $y$ direction or double the forcing wavelength. These are therefore `subharmonic' disturbances as they have half the spatial frequency of the forcing. The blue region shows that the inertial (Newtonian) instability is most unstable when $\mu=0$, when there is no modulation and perturbations have the same wavelength as the forcing, i.e. `harmonic' disturbances. The neutral curve in \cref{floquet}a resembles fig. 1 in \citet{BOFFETTA_CELANI_MAZZINO_PULIAFITO_VERGASSOLA_2005}, which considered the stability of a system equivalent to $n=64$ and $k\in \mathbb{N}/64$. They identified the distinct elastic and inertial instabilities, but our plot allows us to in addition identify the most unstable wavelength of perturbation. It is generally the subharmonic that determines the linear stability of the centre-mode. The neutral curves of \cref{floquet}b are zoomed out and use a log scaling which reveals that there is a part of the parameter space where even lower order harmonics are most unstable.
The $\mu=0$ inertial instability is shown in \cref{floquet}c. Its trace field is antisymmetric about lines of peak base velocity, and it has zero frequency. The $\mu=1/2$ centre-mode is shown in \cref{floquet}d, and is clearly a modulated version of the $\mu=0$ centre-mode shown in \cref{Re-WE_neutral_curves}d). Similarly, we plot the preferred mode when lower order harmonics of the centre-mode are most unstable in \cref{floquet}e. This suggests that all elastic instabilities are centre-modes for all Floquet modes, not just when $\mu=0$.


The dependence of the centre-mode growth rate on $\mu$ is shown in  \cref{floquet}f. We consider vanishing inertia ($Re=0$), and consider the maximum growth rate across all wavenumbers, $\sigma^* \coloneqq max_{k \in \mathbb{R}} \sigma$, as $W$ varies for fixed $\mu$ ($k^*$ is defined as  the most unstable wavenumber). We see that the harmonic disturbances ($\mu=0$) are the most stable of those plotted. The subharmonic with modulation $\mu=1/2$ is the most unstable, and then subsequent modulations increase in stability. 

%
%

%
%
\subsection{The inertialess centre-mode in the upper-convected Maxwell fluid}\label{UCM}

So far we have mainly limited our results to the specific choice of $\beta=0.95$. At this dilute concentration, the elastic instabilities in vKf for both $n=1$ and $n>1$ have been identified as the centre-mode. We now demonstrate that the centre-mode exists not only when $\beta \sim 1$, but for all $\beta$. It even exists when $\beta=0$ and the model reduces to the UCM fluid. 

\begin{figure} 
\centering
\includegraphics[width=0.85\textwidth]{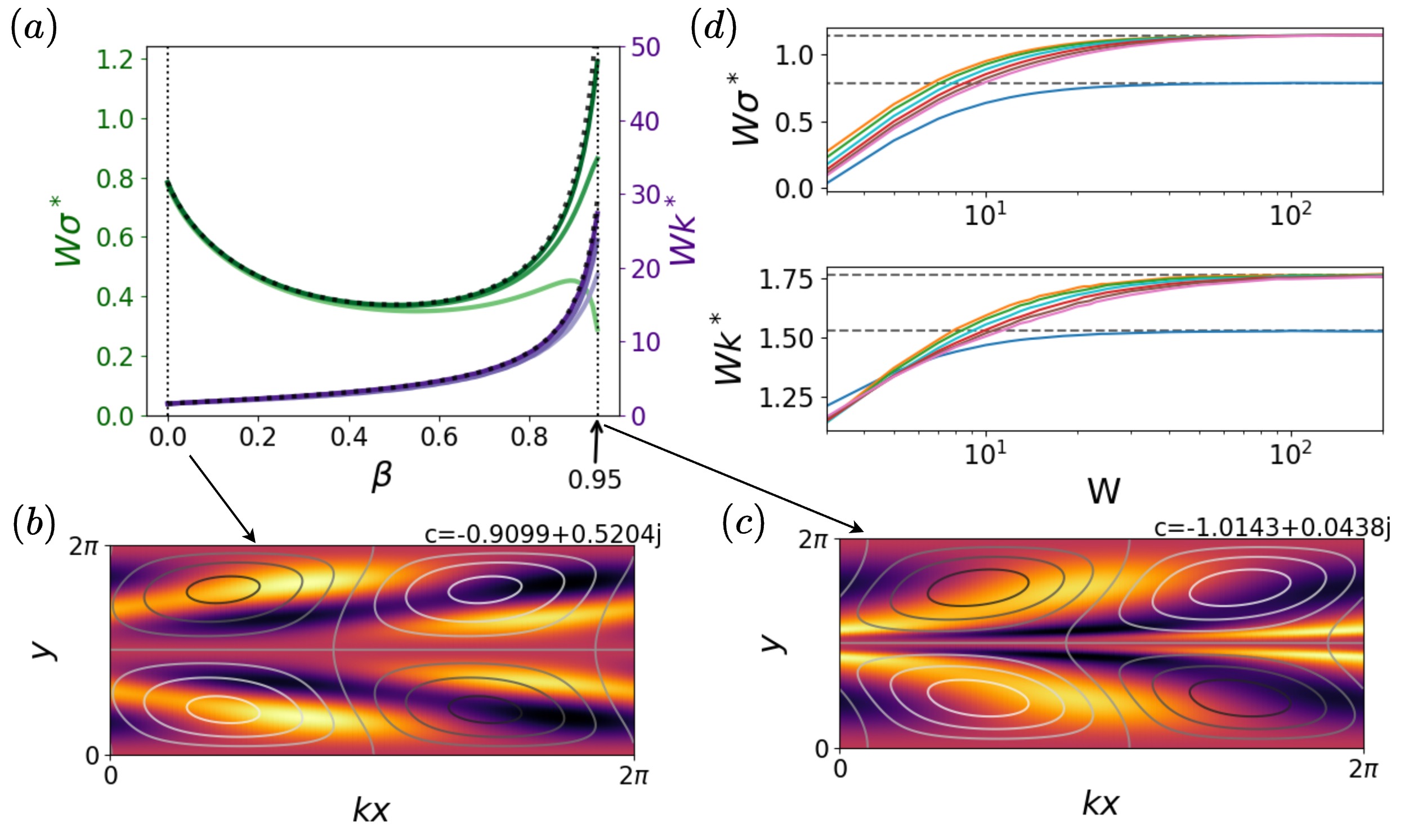}
\caption{$a)$ The maximum growth rate $\sigma^*$ and most unstable wavenumber $k^*$ as $\beta$ varies when $Re=0$, $\varepsilon=0$, $\mu=0$ and  $W=40, 80, 160$ (light to dark). Asymptotics derived in \cref{asymptotic_W} are shown by the black dotted lines. Most unstable eigenfunctions are shown for $W=160$ and $b)$ $\beta=0$, $c)$ $\beta=0.95$ with colours showing the polymer stress trace field and contours showing the streamfunction. $d)$ $\sigma^*$ and $k^*$ in the inertialess UCM fluid for various Floquet modes with $\beta=0$, $Re=0$, $\varepsilon=0$ and $\mu=0, 1/2, 1/3, ..., 1/7$ with colours as in \cref{floquet}. The asymptotic limits as $W\rightarrow\infty$ are shown by horizontal black dashed lines. When $\mu=0$, $W\sigma^* \rightarrow 0.784$ and $Wk^* \rightarrow 1.526$, while when $\mu>0$, $W\sigma^* \rightarrow 1.139$ and $Wk^* \rightarrow 1.764$. The centre-mode is therefore generic across $\beta$, existing even in the UCM fluid, and $k^* \sim W^{-1}$ and $\sigma^* \sim W^{-1}$.}\label{UCM_CM}
\end{figure}

We plot in \cref{UCM_CM}a the maximum growth rate $\sigma^*$ and the most unstable wavenumber $k^*$ of the inertialess instability as $\beta$ varies for the $n=1$ system. This shows that there is a smooth continuation of the centre-mode at $\beta=0.95$ to the instability seen at lower $\beta$, suggesting that the instability seen at low $\beta$ is also the centre-mode. Eigenfunctions are plotted at both $\beta=0$ and $\beta=0.95$ in figures \ref{UCM_CM}b and \ref{UCM_CM}c and clearly resemble each other, again suggesting that the instability at $\beta=0$ is the centre-mode. The centre-mode is not suppressed by low $\beta$ in vKf.

Figure \ref{UCM_CM}d shows the behaviour of the various Floquet modes in the inertialess UCM fluid. As $W\rightarrow\infty$, all harmonics tend to the same growth rate that is more unstable than the $\mu=0$ mode. Of these harmonics, the $\mu=1/2$ mode is the most unstable, as was the case when $\beta=0.95$ with vanishing inertia (shown in figure \ref{floquet}g). The subharmonic is therefore most unstable for both high and low $\beta$.

These plots also demonstrate that the correct asymptotic scalings for the inertialess centre-mode are $k^* \sim W^{-1}$ and $\sigma^* \sim W^{-1}$ as $W\rightarrow \infty$. This is shown across all $\beta$ when $\mu=0$ (see \cref{UCM_CM}a) and various $\mu$ when $\beta=0$ (see \cref{UCM_CM}d). The equations governing the inertialess asymptotic limit of $W \rightarrow \infty$ are identified in \cref{asymptotic_W}, and these asymptotics are plotted in both \cref{UCM_CM}a and \cref{UCM_CM}d, showing their validity. The centre-mode is therefore present in a very simple fluid when $Re=\beta=\varepsilon=0$ and $W \rightarrow \infty$, demonstrating that while elasticity is required for the instability to exist, inertia, viscosity and polymer diffusion are not.

%
%

%
%
\subsection{Relaminarisation in the $W\rightarrow \infty$ limit}\label{relaminarisation}

The flow becomes linearly stable as $W\rightarrow \infty$ for any fixed domain length. This is due to the centre-mode only being unstable to a pocket of wavenumbers that scale like $k \sim W^{-1}$, as suggested by the asymptotic analysis performed in \cref{asymptotic_W}. Hence, at sufficiently large $W$, all unstable wavelengths are longer than the channel itself, meaning the system is not susceptible to the centre-mode instability.

\begin{figure} 
\centering
\includegraphics[width=0.6\textwidth]{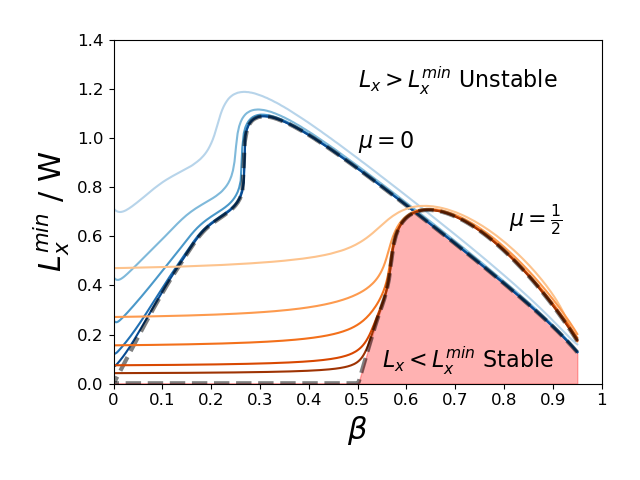}
\caption{The minimum $L_x$ at which laminar flow becomes unstable to perturbations with $\mu=0$ (blue) or $\mu=1/2$ (orange), when $Re=0$, $\varepsilon=0$ and $W=50, 100, 200, 500, 1000$ (light to dark). Black dashed lines correspond to the asymptotic limit described in \cref{asymptotic_W}, and they cross over at $\beta=0.62$. The only region which is stable as $W\rightarrow \infty$ is shaded in red. This confirms that as $W \rightarrow \infty$, $L_x^{min} \sim W$, meaning only very long channels are linearly unstable for large $W$. For $\beta<0.62$, the $\mu=0$ harmonic stabilises before the $\mu=1/2$ subharmonic as $L_x/W\rightarrow 0$, while the opposite is true when $\beta>0.62$.}  \label{relaminar}
\end{figure}

It will be useful to define $L_x^{min} \coloneqq 2\pi / k_{max}$, which is the shortest domain in which instability can be found, i.e. the system is stable when $L_x < L_x^{min}$. Figure \ref{relaminar} demonstrates that $L_x^{min} \sim W$, and hence that as $W \rightarrow \infty$, any finite $L_x$ eventually becomes stable. 
Figure \ref{relaminar} also shows that the order in which the Floquet modes stabilise as $L_x/W \rightarrow 0$ depends on the concentration $\beta$. This limit is achieved for fixed $L_x$ as $W \rightarrow \infty$. In this limit, when $\beta>0.62$ the $\mu=1/2$ mode stabilises before the $\mu=0$ mode, while the opposite is true for $\beta < 0.62$. This is consistent with \cref{floquet}f in which $\beta=0.95$, and the $\mu=1/2$ mode stabilises before the $\mu=0$ mode for negligible inertia as $W\rightarrow \infty$.

%
%

%
%
\section{Subcriticality and Exact Coherent Structures} \label{subcriticality}

Subcritical behaviour can be seen in the centre-mode in a channel \citep{Buza2022b, Buza2022a} and a pipe \citep{Dongdong2021}. In this section we demonstrate that the same is also true in Kolmogorov flow. We identify the structures on the upper branch for both $n=1$ and $n=2$, and find other stable exact coherent structures on a number of solution branches. The presence of an elastic travelling wave was first described by \citet{Berti2010} in a system equivalent to $L_x=L_y=8\pi$ (i.e. $n=4$), and we expand upon this by identifying a number of distinct elastic waves and equilibria in a simpler system with $L_x=L_y=4\pi$ (i.e. $n=2$). No chaotic behaviour was identified here, and so while our choice of parameters produces a number of different solutions, it is simple to track the solutions as we change $W$. We will later increase $L_x$ from this value, which allows the system to become chaotic.

\begin{figure} 
\centering
\includegraphics[width=0.8\textwidth]{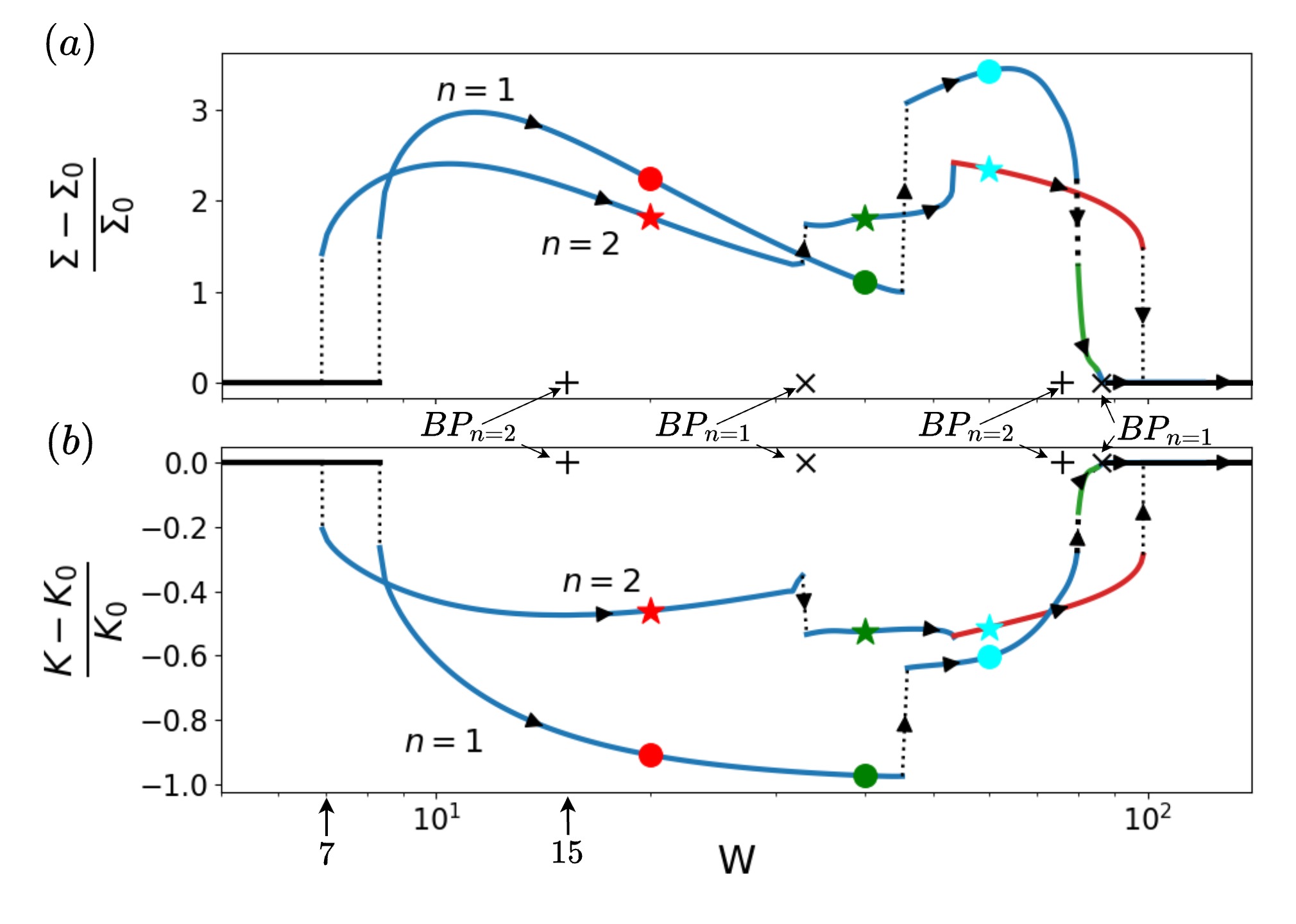}
\caption{Bifurcation plots for $\beta=0.95$, $Re=0.5$, $\varepsilon=10^{-3}$ and $L_x=4\pi$. These show the deviation of the volume averaged $a)$ trace $\Sigma$ and $b)$ kinetic energy $K$ from the laminar state, which has trace and kinetic energy $\Sigma_0$ and $K_0$. We show stable solutions in both the $n=2$ system and the $n=1$ system. Blue corresponds to travelling wave solutions, red to equilibria and green to limit cycles. The polymer stress trace of the solution at each of the 6 symbols are shown in \cref{ECS_plots}. Bifurcation points (BP) due to the linear instability are shown with black crosses at $W=33, 86$ when $n=1$ and black pluses at $W=15, 76$ when $n=2$.}  \label{bifurcation}
\end{figure}

We consider the bifurcation plot of the centre-mode instability in \cref{bifurcation}. To produce this plot we begin with laminar flow at a value of $W$ that is linearly unstable, and then add white noise and allow the system to reach its stable final state. $W$ is then adiabatically decreased until a saddle-node is identified. From this saddle-node, $W$ is then increased adiabatically, and the resultant stable branch of the bifurcation diagram is plotted in \cref{bifurcation}. This procedure was followed for both $n=1$ and $n=2$. The solutions on the $n=1$ branch can be used to construct a solution branch when $n=2$ by repeating all fields twice in the y direction, however the stability of this constructed branch may be different in the $n=2$ system. In fact, for $W=20,40, 60$ the solutions constructed from the $n=1$ branch are unstable in the $n=2$ system (not shown). We plot the mean kinetic energy $K = \langle|\vb u|^2 \rangle_V/ 2$ and the mean trace $\Sigma =  \langle \tau_{xx} + \tau_{yy} \rangle_V$ of the solution branches, where $\langle \cdot \rangle_V$ denotes an average over the domain. Both metrics are normalised using their values in the laminar flow, $\Sigma_0$ and $K_0$. This bifurcation plot consists of a number of different branches, all of which are either laminar (black), travelling waves (blue), equilibria (red) or periodic orbits (green). For the chosen parameters of $L_x=4\pi$ with $n=1$ or $n=2$, no chaotic behaviour was identified.

%
%

\begin{figure} 
\centering
\includegraphics[width=0.7\textwidth]{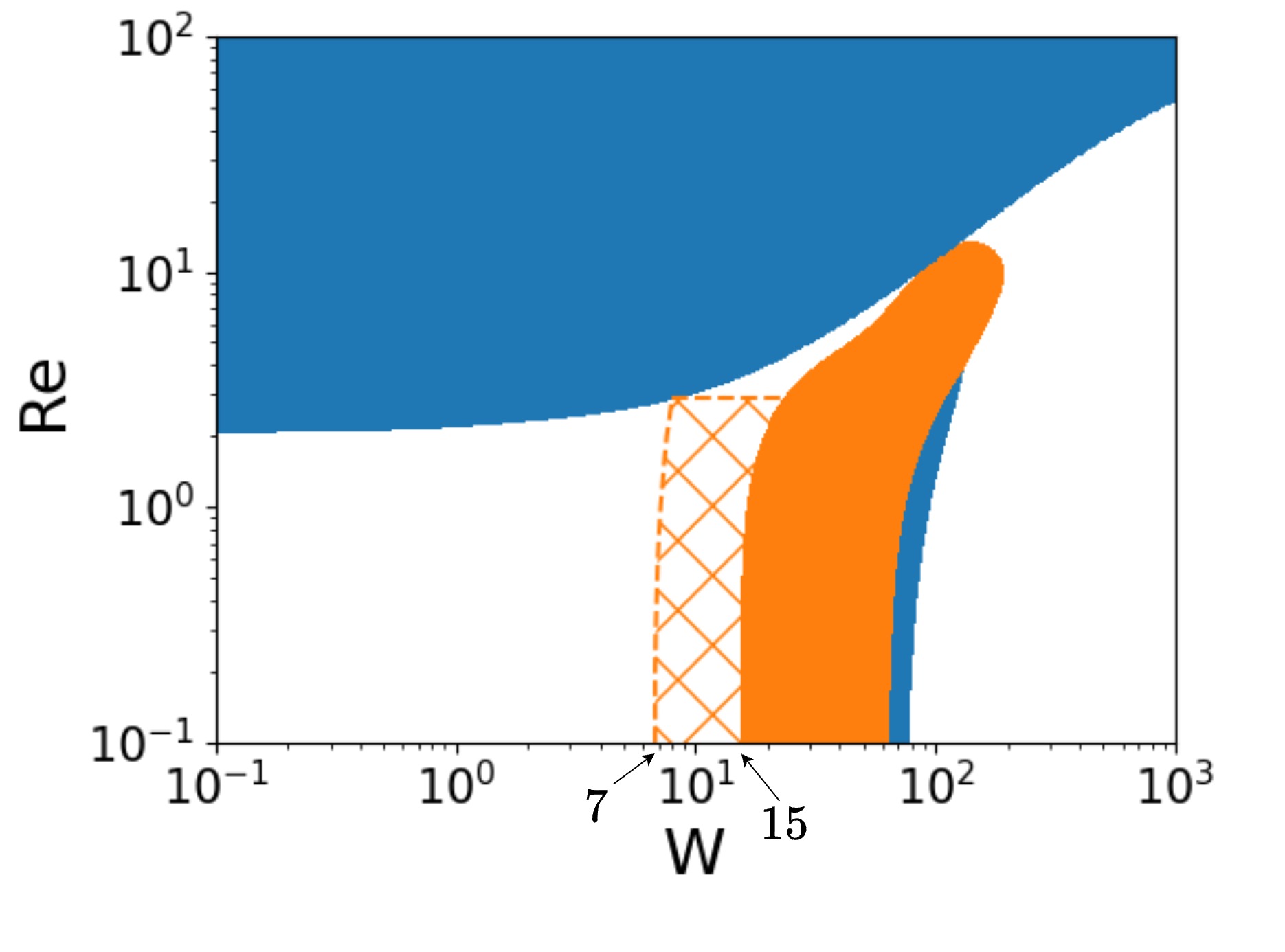}
\caption{The most unstable Floquet modes in the $(Re, W)$ plane for $\beta=0.95$, $\varepsilon=10^{-3}$, with instabilities over wavenumbers $k \in \mathbb{N}/2$ considered and Floquet modes $\mu=0$ (blue) and $\mu=1/2$ (orange). The solid colours therefore show the linear stability of a simulation with $n=2$ and $L_x=4\pi$. The orange hatched region corresponds to areas in which the centre-mode in this geometry has been shown to be subcritical. For $Re<<1$, we see that the system is unstable to finite amplitude perturbations at $W\approx 7$ but to infinitesimal perturbations at $W\approx 15$}  \label{floquet_discrete}
\end{figure}

This plot shows that there is a saddle-node bifurcation marking the smallest $W$ at which the system is unstable to finite amplitude disturbances, demonstrating subcriticality. We illustrate in \cref{floquet_discrete} how this impacts the stability of vKf by plotting a neutral curve for the system that shows regions of linear instability, as well as a hatched region corresponding to where the system is subcritical. This shows the elastic system becomes unstable to finite amplitude perturbations at $W \approx 7$, and becomes linearly unstable at $W \approx 15$. 
The bifurcation plot in \cref{bifurcation} shows that at large $W$, the solutions eventually merge with the laminar state, rather than sustaining any instability. This is consistent with the relaminarisation discussed in \cref{relaminarisation}, which showed how the laminar state is linearly stable for sufficiently large $W$.

%
%



\begin{figure} 
\centering
\includegraphics[width=0.6\textwidth]{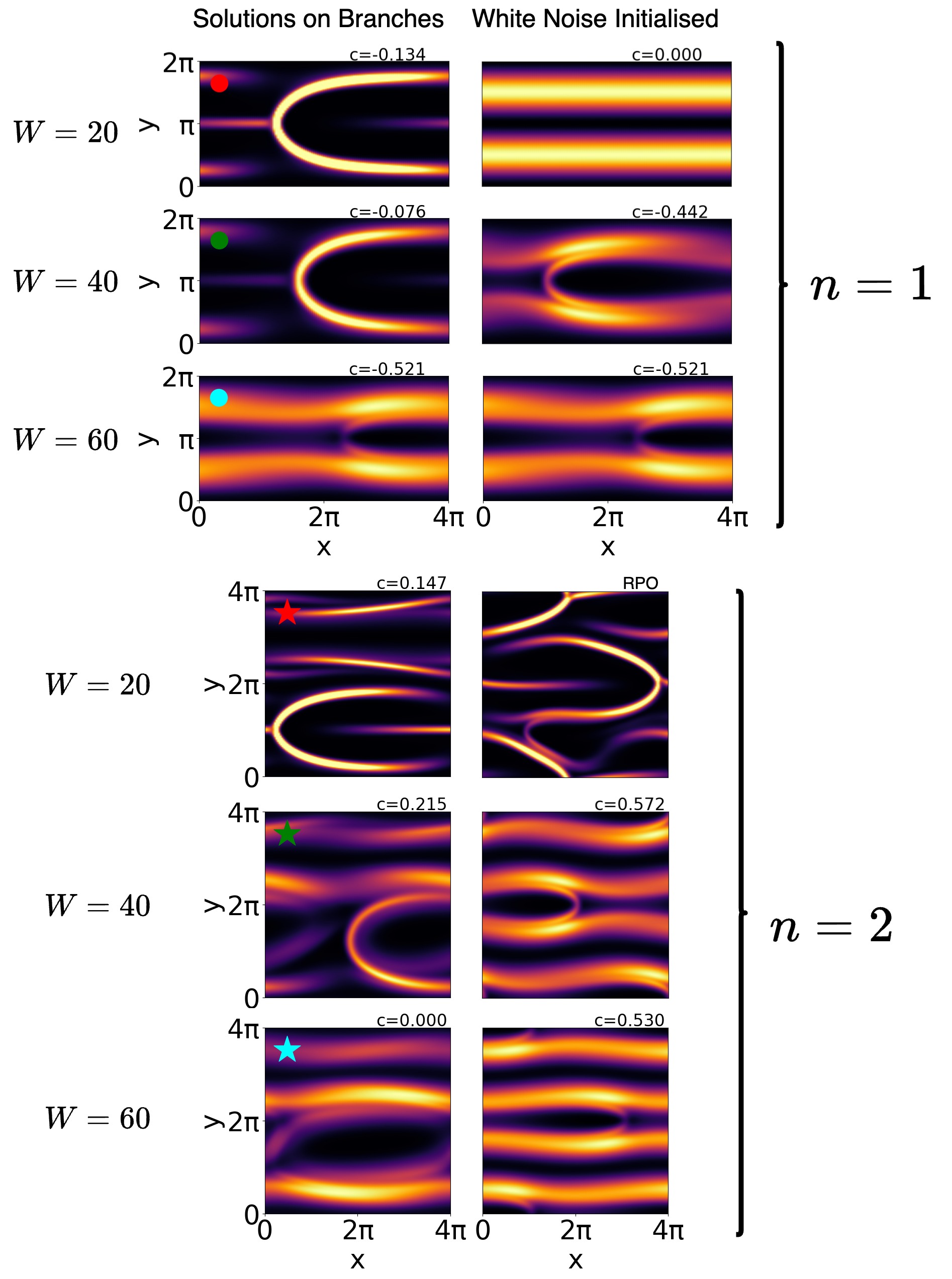}
\caption{The trace $T_{xx} + T_{yy}$ of final states for $\beta=0.95$, $Re=0.5$, $\varepsilon=10^{-3}$ and $n=1,2$. The left column corresponds to solutions on the branches shown in the bifurcation plot in \cref{bifurcation}, while the right shows final states reached when the system is initialised with laminar flow and low amplitude white noise. The wave speeds $c$ of all but one solution is shown, identifying which states are equilibria or travelling waves. For $n=2$, $W=20$ the white noise initialised solution is an RPO, and so has no wave speed.}\label{ECS_plots}
\end{figure}

The trace fields of these solution branches at $W=20,40,60$ (all marked by symbols on the bifurcation diagram) are shown on the left of \cref{ECS_plots}. While these solution branches all represent an attractor of the system, we note that they are not generically the final state for any initial condition. We plot the final states reached when the system is initialised with laminar flow and low amplitude white noise on the right of \cref{ECS_plots}. This protocol produced final states that were different to the solutions found in the bifurcation plot at $W=20,40,60$ for $n=2$ and at $W=20,40$ for $n=1$. This suggests that the basin of attraction for the identified branches in \cref{bifurcation} is small. In all cases, a resolution of at least $64$ modes per $2\pi$ in either $x$ or $y$ direction was  used. Doubling the resolution in each direction did not qualitatively change any of the final states of \cref{ECS_plots}.

Many of the identified exact coherent structures share the arrowhead as a common feature, as was the case with the elastic wave found in \citet{Berti2010}. 
Generally this arrowhead moves in the direction in which the arrow points, however this is not always true  - the travelling waves on the solution branch seen when $n=2$ for $W=20$ and $W=40$ move in the opposite direction to what one might expect.

%
%

%
%
\section{Elastic Turbulence}\label{turbulence}

So far we have examined a number of solutions in the Kolmogorov system when $L_x=L_y=4\pi$, none of which were chaotic. We now change our domain length $L_x$ to see how this can introduce chaos into the system. Elastic turbulence has been identified in vKf in systems equivalent to $L_y=8\pi$ \citep{Berti2008}, and on increasing $W$ the final state can transition from a travelling wave to a periodic state and then onto something chaotic \citep{Berti2010}. \citet{Berti2008}  comment on the presence of `wavy patterns' in the turbulence that are made up of `filamental structures'. Our results indicate that the centre-mode instability is responsible for this turbulence with these structures corresponding to unstable arrowheads. 

%
%
\subsection{The centre-mode is responsible for transition to turbulence}\label{centre_mode_responsible}

Motivated by the presence of the centre-mode arrowhead in many of the coherent structures in \cref{subcriticality}, we now demonstrate that the centre-mode instability is the cause of turbulence in this system. In fact, in the absence of any other elastic instabilities, all the dynamics ultimately comes from the centre-mode instability.

\begin{figure} 
\centering
\includegraphics[width=0.85\textwidth]{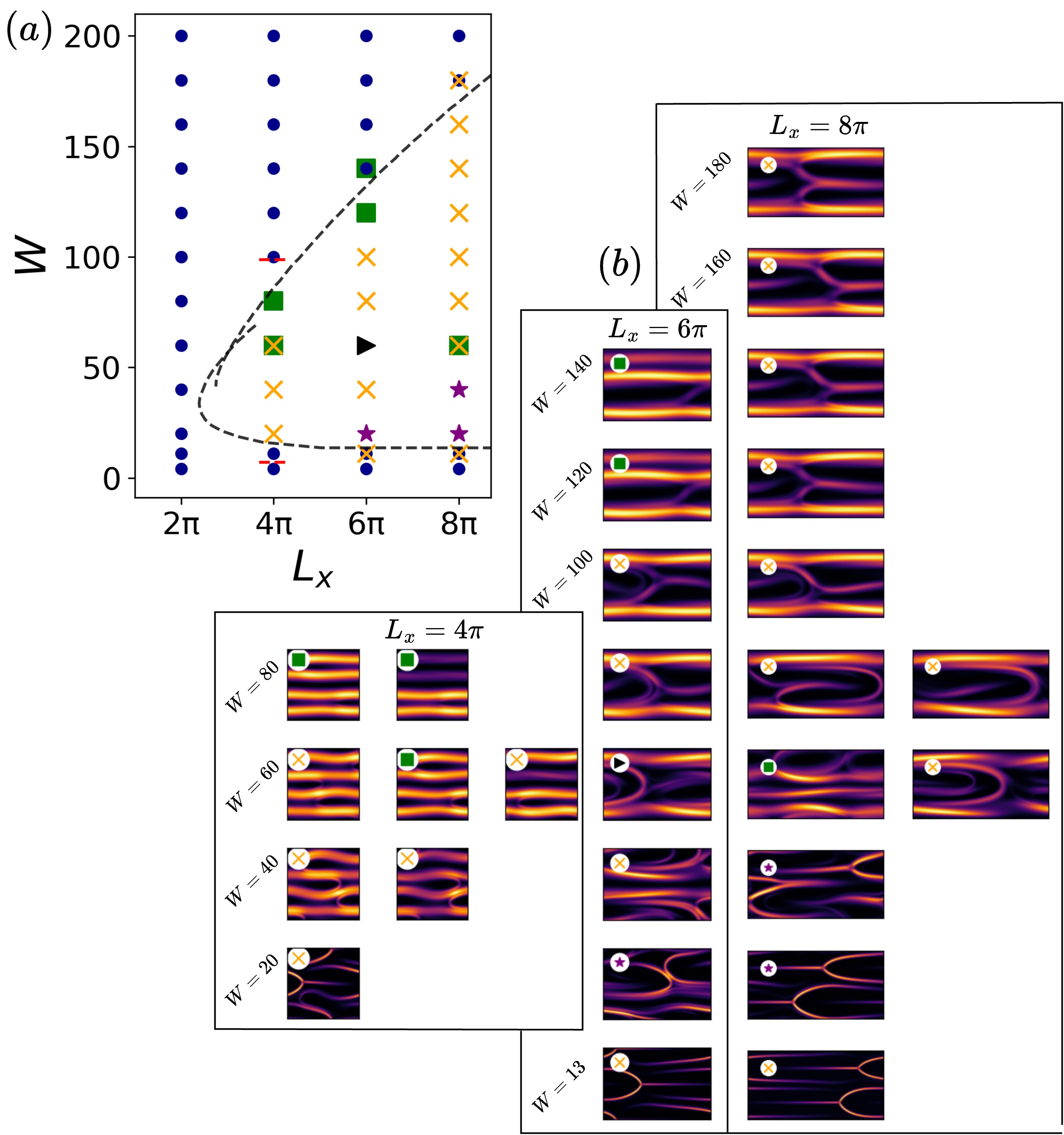}
\caption{$a)$ State diagram of the final states identified when $\beta=0.95$, $Re=0.5$, $\varepsilon=10^{-3}$ and $L_y=4\pi$. For each parameter setting, we simulated the fluid initialised with random finite amplitude disturbances (see main text for the protocol). States seen are laminar (blue circle), travelling waves (green square), periodic orbit (yellow cross), quasi-periodic orbit (black triangle) and chaos (purple star). Dashed black line shows the smallest $L_x$ at which linear instability exists. On the low shear boundary, it is the $\mu=1/2$ Floquet mode that is marginally linearly unstable, while on the high shear boundary it is the $\mu=0$ mode. Dashed red line shows the boundary of where the flow is known to be subcritical when $L_x=4 \pi $ as per \cref{bifurcation}. $(b)$ All distinct non-laminar final states identified, marked with a symbol denoting the type of state as in $(a)$. This plot demonstrates that instabilities have only been identified close to regions of parameter space in which the centre-mode is linearly unstable, and many of these states contain an arrowhead. This suggests all non-laminar states originate from the centre-mode linear instability.}\label{state_diagram}
\end{figure}

We first produce a state diagram in \cref{state_diagram}a, which shows what types of final states are reached when simulations are initialised with finite-amplitude disturbances as the Weissenberg number $W$ and channel length $L_x$ are varied. Each state was found by taking the laminar base flow and adding a random perturbation to the first 50\% of the Fourier modes in fields $u$, $T_{xx}$ and $T_{xy}$, with an amplitude of 10\% of the mean laminar base field. All remaining initial fields were constructed from these. We simulated each pair of parameters ($W$, $L_x$) 5 times, and identify the final states as either laminar (blue circle), a travelling wave (green square), a periodic orbit (yellow cross), a quasi-periodic orbit (black triangle) or chaotic (purple star) using the procedure discussed in \cref{ECS}. In regions of linear stability, this protocol produced exclusively laminar solutions, so to demonstrate subcriticality we simulated an additional 10 runs as well. In these, random perturbations were added to the first 5\% of Fourier modes with an amplitude of 20\% of the laminar base field for 5 runs, and an amplitude of 10\% of the laminar base field was set in the other 5. This focus on longer wavelength perturbations was sufficient to excite subcritical solutions, however not all subcritical solutions were found, as a subcritical travelling wave solution that is known to exist at $W=13, L_x=4\pi$ (see \cref{bifurcation}) was not found here. For some parameters these runs identified multiple types of final states, which are indicated as overlaid symbols in \cref{state_diagram}.
These simulations used a resolution of $(N_x, N_y)=(128, 192)$ Fourier modes with a timestep of $dt=8\times 10^{-2}$ and were confirmed using $(N_x, N_y)=(256, 256)$.

Non-laminar solutions were only seen in regions of parameter space close to where the centre-mode is linearly unstable, suggesting that these states originate from the centre-mode bifurcation. We plot all distinct final states in \cref{state_diagram}b. While some states look similar (e.g. those at $W=80$, $L_x=8\pi$), kinetic energy time series differ (not shown).

For the parameters considered, the shortest channel in which chaotic behaviour was found was $L_x=6\pi$, at $W$ corresponding to the low shear region of where the centre-mode is linearly unstable. To check that the chaos was self-sustaining and not transient, we ran all chaotic simulations for $10^5$ time units of $2\pi U_0 / L_0$, and saw no collapse in the time series of $K$. 


%
%

\begin{figure} 
\centering
\begin{subfigure}[b]{0.7\textwidth} 
\includegraphics[width=\textwidth]{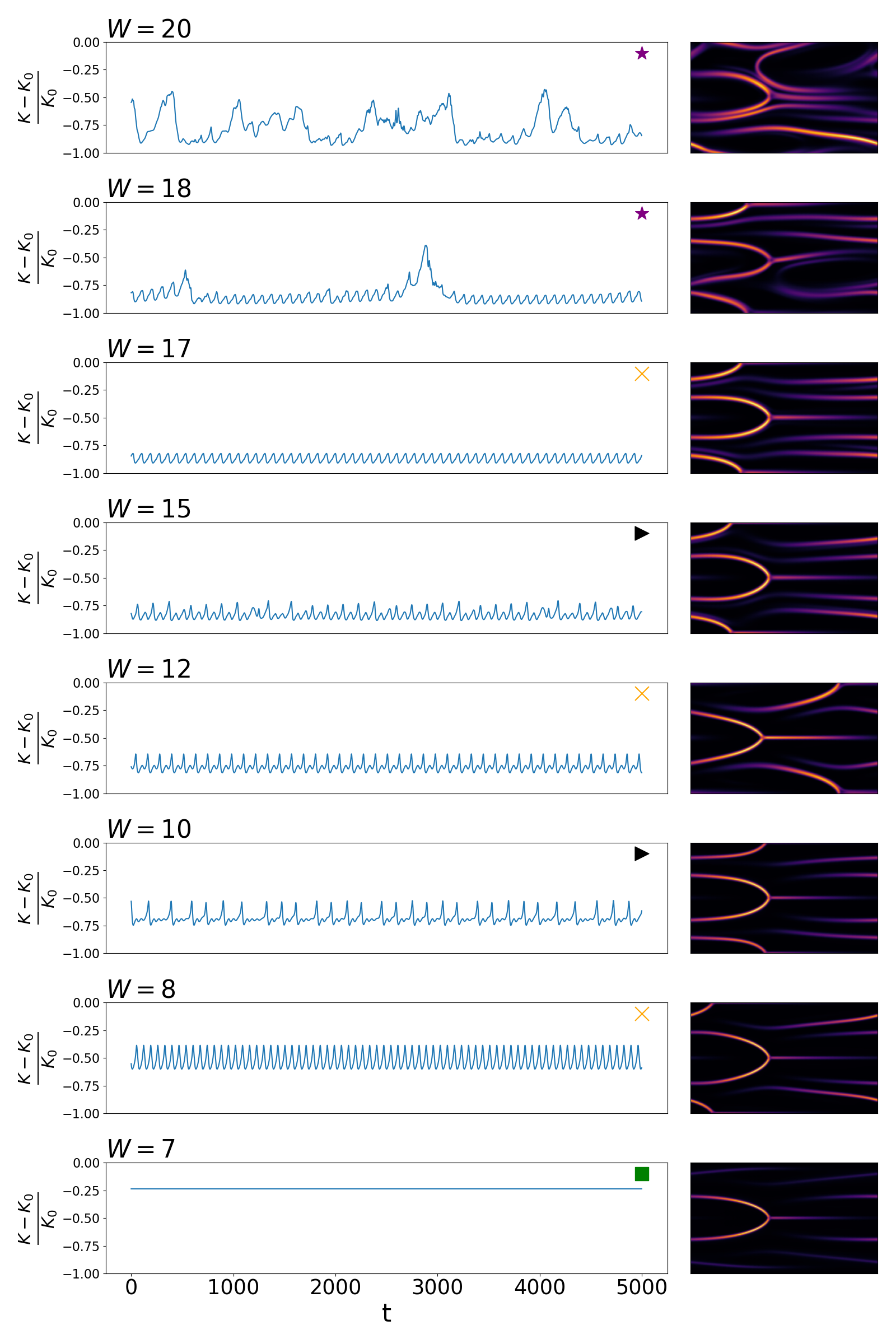}
\end{subfigure}
\caption{Kinetic energy time series of solutions (left), with the trace field at $t=5000$ (right) as $W$ is lowered from $W=20$. Parameters are $\beta=0.95$, $Re=0.5$, $\varepsilon=10^{-3}$, $n=2$ and $L_x=6\pi$. Symbols, as in \cref{state_diagram}a, show chaos (purple star), quasi-periodic orbits (black triangle), periodic orbit (yellow cross) and a travelling wave (green square). This shows that the turbulent state at $W=20$ is connected to states that strongly resemble the centre-mode arrowhead.}\label{chaos_branch_timeseries}
\end{figure}

To examine the transition scenario, the turbulent state at $W=20$ and $L_x=6\pi$ was taken as a starting point  and then $W$ lowered adiabatically.  \cref{chaos_branch_timeseries} shows how the flow transitions from one where the time series of $K$ is chaotic, to one that is periodic via regions of quasi-periodicity as $W$ is decreased, before eventually becoming a constant when the state is a travelling wave. At $W=18$, the onset of turbulence is marked by the presence of bursting events that temporarily increase $K$ seemingly at random, before returning to periodic behaviour. When $W$ is decreased further, periodic behaviour (e.g. at $W=17$) and quasi-periodicity (e.g. at $W=15$) are found, until a travelling wave state is realised at $W=7$. The periodic and travelling wave solutions closely resemble the centre-mode arrowhead.

\begin{figure} 
\centering
\begin{subfigure}[b]{0.8\textwidth} 
\includegraphics[width=\textwidth]{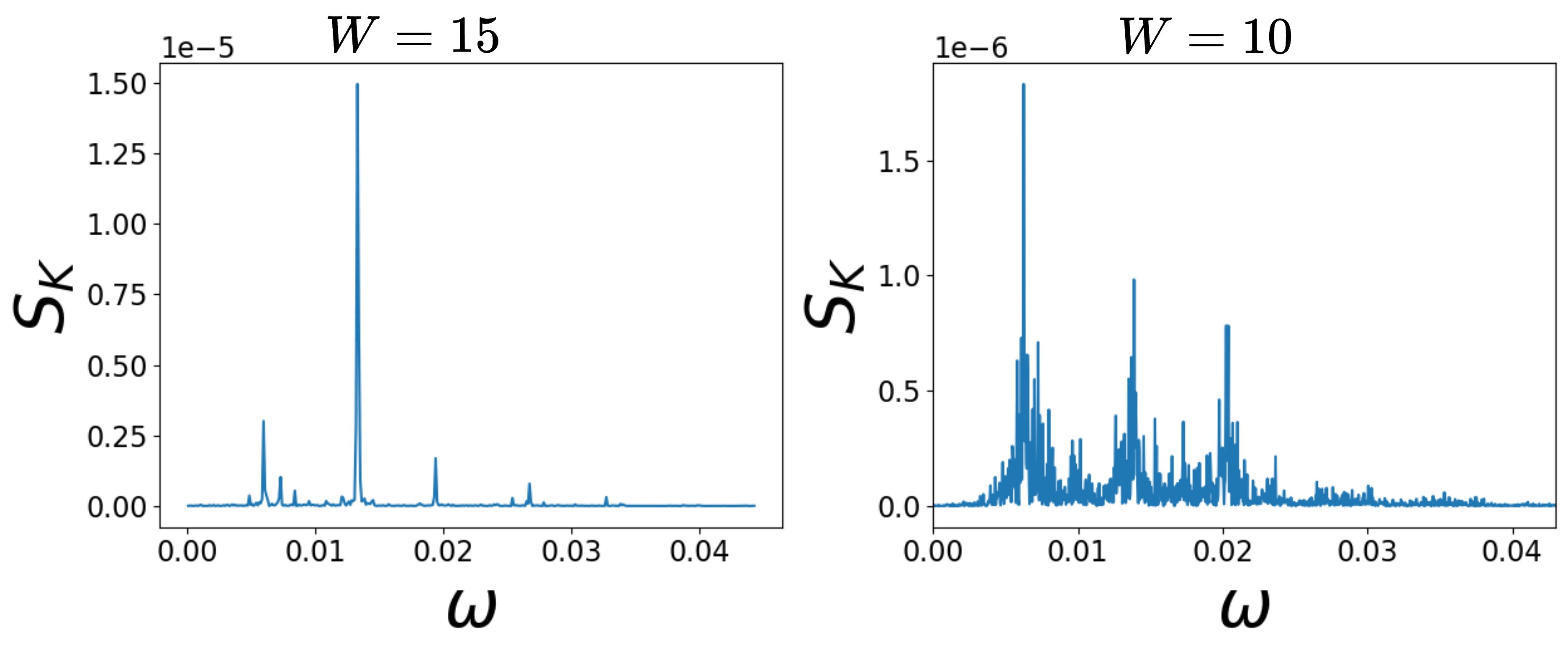}
\end{subfigure}
\caption{Power spectra for the states from \cref{chaos_branch_timeseries} with $W=15$ and $W=10$. This shows that the state at $W=15$ is truly a quasi-periodic orbit with discrete incommensurate frequencies, while the state at $W=10$ has a broadband of frequencies. }\label{qpo_spectra}
\end{figure}

We plot the frequency spectra of the quasi-periodic behaviour at $W=15$ and $W=10$ in \cref{qpo_spectra}, where
\begin{equation}
S_K(\omega) = \left|\frac{1}{T} \int_0^TK(t)e^{-i\omega t} dt\right|^2 \label{S}
\end{equation}
is the Fourier transform of the kinetic energy $K$ over a long time $T=50,000$. While $W=15$ clearly shows discrete incommensurate frequencies, $W=10$ shows a low-level broadband of frequencies. We still describe the latter as quasi-periodic however, as it contains a small number of dominant frequencies that are incommensurate. Fig. \ref{chaos_branch_timeseries} used $\varepsilon=10^{-3}$, and was run with a resolution of $(N_x, N_y) = (196, 128)$. The results were robust to a doubling of the resolution in both directions. In addition, we checked that $\varepsilon=10^{-4}$ ran at $(N_x, N_y) = (384, 384)$ produced qualitatively similar results, despite the arrowheads being thinner: see \cref{transition_lower_eps}. The bursting scenario is still identifiable.\\

Finally, it is worth remarking that the transition scenario identified here is not the only one possible in vKf. In the regime discussed with $n=2$ and $L_x=6\pi$, a bifurcation triggers irregular bursting events that become more frequent with increasing $W$. ET is then seen above a critical $W$ where the disordered state exists at all times. An alternative transition to turbulence has been recently identified when $n=1$ and $L_x=8\pi$, in which a sequence of period doubling bifurcations cascade into turbulence \citep{Nichols2024}.

%
%

%
%

%
%
\subsection{Elastic turbulence properties}\label{properties}

\begin{figure} 
\centering
\includegraphics[width=0.9\textwidth]{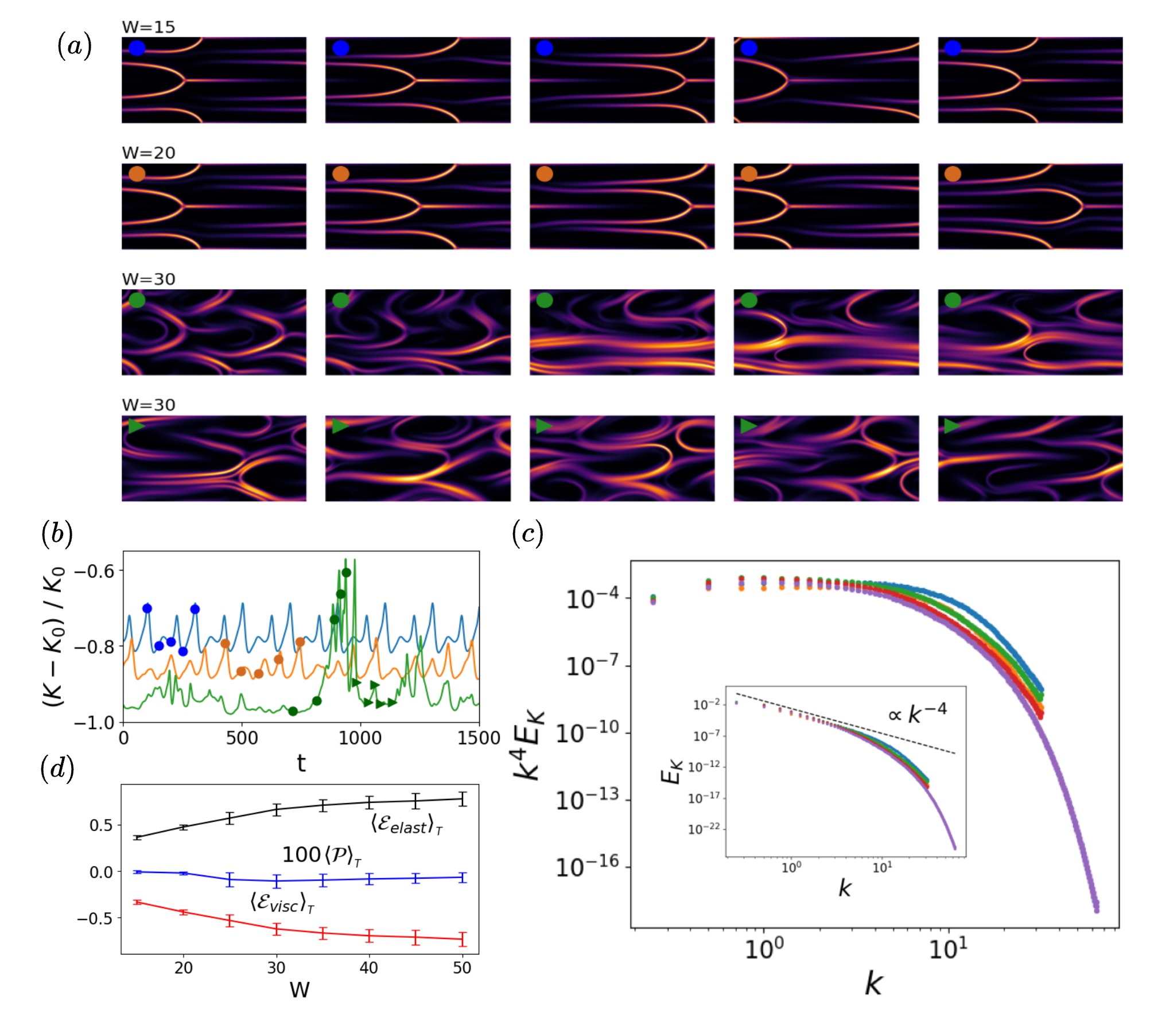}
\caption{$a, b)$ Simulations when $W=15$ (periodic, blue), $W=20$ (low-dimensional chaos, orange) and $W=30$ (turbulent, green), with rows of $a)$ showing the trace field evolving over time, with times marked on the time series of $K$ in $b)$. $c)$ The compensated power spectra for $W=15$ (blue), $20$ (orange), $30$ (green), $40$ (red), $50$ (purple), suggesting a regime where $E_K \sim k^{-4}$. $d)$ Contributions to the kinetic energy due to the base shear ($\mathcal{P}$), elastic forces ($\mathcal{E}_{elast}$) and viscous forces ($\mathcal{E}_{visc}$) as $W$ varies. Each quantity is averaged over a long time ($T=20000$), with error bars showing plus and minus one standard deviation. Other parameters are $\beta=0.95$, $Re=0.5$, $\varepsilon=10^{-3}$, $n=2$ and $L_x=8\pi$ in all plots.} \label{ET_properties}
\end{figure}

An analysis of ET in vKf is now presented in light of the fact that it occurs due to the centre-mode arrowhead losing stability as $W$ increases. We consider the power spectrum, as well as how energy is produced and dissipated in ET in a domain of $L_x=8\pi$ where turbulence exists across a wider range of $W$ (as per \cref{state_diagram}). These are compared to those of the centre-mode arrowhead structure. Three final states are considered: a periodic state at $W=15$, a weakly chaotic state at $W=20$, and a strongly chaotic at $W=30$: see \cref{ET_properties}. The polymer stress trace field is plotted at various times in \cref{ET_properties}a, and the kinetic energy time series $K$ normalised by the laminar kinetic energy $K_0$ in \cref{ET_properties}b. The periodic state at $W=15$ consists of two arrowheads drifting in the same direction (though still moving relative to each other). The low-dimensional chaos at $W=20$ again has two arrowheads drifting in the same direction, never making contact with each other. However, the increase in $W$ allows the `tail' of the arrowhead to lengthen to the extent it now interacts with its `head'. We verify that the dynamics are truly chaotic, rather than quasi-periodic using the classification process discussed in \cref{ECS}. The chaotic state at $W=30$ is more complicated, but we note two processes can be identified, and we show these in the lower two rows of \cref{ET_properties}a. The third row shows how multiple arrowheads can join together to form a zonal shear that spans the entire streamwise direction. This is a bursting event in which the kinetic energy peaks. The zonal shear then collapses, splitting into multiple arrowheads. The fourth row shows how the collision, coalescence and subsequent splitting of arrowheads contributes to the turbulent flow.

In \cref{ET_properties}c the compensated power spectra of ET in vKf is plotted for various $W$ which is consistent with $E\sim k^{-4}$. This is the scaling found by \cite{Rota2023, Lellep2024} and \cite{Singh2023}, and is close to that of \cite{Berti2010} who found $E\sim k^{-3.8}$. Interestingly
the periodic state at $W=15$, the weakly chaotic state at $W=20$, and the strongly chaotic states at $W=30$ all show similar scaling laws, suggesting that the centre-mode arrowhead contains the same small scale structures present in elastic turbulence. Each of the simulations was run at a resolution of $(N_x, N_y) = (256, 256)$, except for $W=50$ where $(N_x, N_y) = (512, 512)$ was used. This high resolution run gave a power spectrum that was visually similar to one run at the lower resolution (not shown).

We now consider the total energy balance and see how the inputted energy due to the base shear compares to the viscous and elastic dissipation. The perturbative kinetic energy budget is derived by expanding the momentum \cref{governing1} about the base state and dotting this with the velocity perturbation $\vb u'$. This produces
\begin{equation}\label{energy0}
    Re \left(\frac{\partial u'_i}{\partial t} + u'_j \partial_j U_i + U_j \partial_j u'_i + u'_j \partial_j u'_i \right) u'_i = \left(- \partial_i p' +  (1-\beta) \partial_j \tau'_{ji} + \beta \nabla^2 u'_i \right) u_i' 
\end{equation}
where dashed quantities denote real perturbations from the laminar state, and Einstein summation is used over repeated indices. Incompressibility and the product rule allows us to rewrite this as
\begin{equation}
\begin{gathered}\label{energy1}
    \frac{1}{2}\frac{\partial (u'_i u'_i)}{\partial t} + u'_i u'_j \partial_j U_i + \frac{1}{2} \partial_j ( U_j u'_i u'_i) + \frac{1}{2}  \partial_j (u'_j u'_i u'_i)   = - \frac{1}{Re}\partial_i (p' u_i') +  \frac{(1-\beta)}{Re} \partial_j (\tau'_{ji} u_i') \\ -  \frac{(1-\beta)}{Re}  \tau'_{ji}\partial_j u_i' + \frac{\beta}{Re} \partial_j (u_i' \partial_j u'_i) - \frac{\beta}{Re} \partial_j u_i' \partial_j u'_i .
\end{gathered}
\end{equation}
Taking a volume average of \cref{energy1} produces an energy budget of
\begin{equation}\label{energy_budget}
\frac{d}{dt}(KE) =  \mathcal{P} + \mathcal{E}_{elast} + \mathcal{E}_{visc}
\end{equation}
where
\begin{equation}
KE = \frac{1}{2} \langle  \u \cdot \u  \rangle_V, \qquad \mathcal{P} = -\langle \vb \nabla \U : \u \u \rangle_V
\end{equation}
\begin{equation}
\mathcal{E}_{elast} = -\frac{1-\beta}{Re}\bigl\langle \boldsymbol {\tau'} :{\vb \nabla \vb{u'}} \bigr\rangle_V,
\qquad
\mathcal{E}_{visc} = -\frac{\beta}{Re}\bigl\langle {\nabla \vb u'} : \nabla \u  \bigr\rangle_V
\end{equation}
and  $\langle \,\cdot\, \rangle_V$ denoting a volume average.
This energy budget demonstrates that the kinetic energy can be energised or dissipated via the base shear through $\mathcal{P}$, elastic forces through $\mathcal{E}_{elast}$ or viscous forces through $\mathcal{E}_{visc}$. We plot in \cref{ET_properties}d the long time averages of each energising and dissipative term as $W$ varies, noting that integrating \cref{energy_budget} over a long time $T$ gives $\langle\mathcal{P}\rangle_T + \langle\mathcal{E}_{elast}\rangle_T + \langle\mathcal{E}_{visc}\rangle_T = 0$. We see that the only energising process is elasticity, that the base shear contributes little to the kinetic energy, and that viscous forces dissipate energy. This is consistent with \cite{Buza2022a}, which considers the energy budget in channel flow of the centre-mode eigenfunctions at higher $Re$. We see that the energy budget is qualitatively similar for flows which are chaotic ($W=30$) and for those which are periodic ($W=15$), but that as $W$ increases, elasticity provides more energy to the flow which is dissipated via viscosity.

%
%
\section{Discussion}\label{conclusion}

%
%
In this paper we have found that the only instability operative in 2D Kolmogorov flow of an Oldroyd-B fluid at vanishing Reynolds number is the centre-mode instability (e.g. PDI does not occur here). As a result, we can confirm that the instability found by \citet{BOFFETTA_CELANI_MAZZINO_PULIAFITO_VERGASSOLA_2005} at very low $Re$ in Kolmogorov flow is  the same instability found at much larger $Re$ and $W$ in pipe flow over a decade later \citep{Garg_Chaudry_Khalid_Shankar_Subramanian_2018}.
The fact that this instability only exists for $Re \gtrsim 63$ in the pipe flow of an Oldroyd-B fluid \citep{Chaudhary_Garg_Subramanian_Shankar_2021}
suggested it was an elasto-inertial instability. It was later shown to be purely elastic in origin however, by tracking the instability down to $Re=0$ in channel flow (using the extreme parameter values $\beta> 0.9905$ and $W >974$:  \cite{Khalid_Shankar_Subrmanian_2021}), and 
by considering the energy budget \citep{Buza2022a}. We formally connected the centre-mode instability of the channel and pipe to inertialess Kolmogorov flow, tying together the different flow geometries. The absence of boundaries in Kolmogorov flow is undoubtedly responsible for the substantial downward shift in parameter values, as well as allowing the centre-mode instability to exist at all concentrations $\beta \in [0,1)$ and with different wavelengths to the forcing. Floquet analysis reveals that the subharmonic instability (where the instability has a wavelength twice that of the forcing) is usually the most unstable.

Numerical computations confirm that the centre-mode instability is subcritical with respect to $W$, as it is in channel flow \citep{Page2020, Buza2022b} and pipe flow \citep{Dongdong2021}. The complexity of the nonlinear solutions found increases with the domain length $L_x$, the number of forcing wavelengths $n$ and $W$ but all are built upon  arrowhead structures as found earlier by \cite{Berti2008} and \cite{Berti2010}. Chaotic behaviour occurs only when the domain is long enough ($L_x$ large enough) with the transition scenario depending on how many forcing wavelengths fit into the domain (i.e. $n$). Using a domain which allows two forcing wavelengths ($n=2$), we identified an irregular bursting scenario after the centre-mode arrowhead structure becomes unstable while a concurrent study finds  a period doubling cascade just including one forcing wavelength ($n=1$) \citep{Nichols2024}. In $n>1$, the final chaotic or ET state consists of centre-mode arrowheads colliding, coalescing and then splitting up, with a zonal shear forming during bursting events. Power spectra of this chaos shows a part of the kinetic energy having a $k^{-4}$ spectrum which is already possessed by the centre-mode arrowhead structure.


The sequential `supercritical' transition scenarios seen here and in \cite{Nichols2024} for 2D Kolmogorov flow contrast with current observations in channel flow. There, in 2D, the arrowheads are stable and so do not sequentially breakdown to ET \citep{Lellep2023} or indeed EIT at higher $Re$ \citep{Beneitez_Page_Dubief_Kerswell_2024}. In 3D however, there are instabilities which seem to lead to a weak chaotic state \citep{Lellep2023,Lellep2024}.  Surely there is a counterpart of this in $n=1$ 3D Kolmogorov flow which can be explored with far greater ease given the triply periodic boundary conditions. In turn, investigating why 2D ET can occur in Kolmogorov flow and not obviously in channel flow presents another interesting challenge. In other words, the establishment of a firm instability connnection between boundaried and unboundaried viscoelastic straight shear flows can help boost the understanding in either. We hope to report on such progress in the near future.

\FloatBarrier 

\newpage

\appendix

%
%
\section{$W \rightarrow \infty$ Asymptotics for the Centre-Mode in Inertialess ($Re=0$) Kolmogorov Flow} \label{asymptotic_W}

We saw in \cref{linear_instability} that Kolmogorov flow is linearly unstable to the centre mode even when inertia is negligible, and so, motivated by this, we consider the asymptotics when $W \rightarrow \infty$, $Re=0$ and $\varepsilon=0$ at fixed $\beta$. This limit is different to the `ultra-dilute' distinguished limit in which $W\rightarrow \infty$ and $\beta \rightarrow 1$ such that $W(1-\beta)$ is finite which is required for the centre mode instability to survive in inertialess channel flow \citep{Khalid_Shankar_Subrmanian_2021, Kerswell_2023}.



To motivate our scalings we re-examine the collapse of the neutral curves in the $(Re, k)$ plane when $W\gg1$ and $Re\ll1$. This is seen in the $E \gg 1$ main loop of neutral curves of \cref{Re-k_neutral_curves}d, in regions where $ERe \gg 1$. Defining $k_{max}$ to be the largest wavenumber that is unstable for a set $W$ and $Re$ in this region, we have $k_{max} \sim W^{-1}$. Motivated by these long wavelength instabilities, we seek rescalings of
$$(k, c, u', v', \tau_{xx}', \tau_{xy}', \tau_{yy}', D) = (\frac{\hat k}{W}, \hat c, \hat u, \frac{\hat v}{W}, W\hat \tau_{xx}, \hat \tau_{xy}, \frac{\hat \tau_{yy}}{W}, \hat D).$$
These factors are determined via considering a dominant balance between the $D\tau_{xx}'$, $D^2 \tau_{xy}'$ and $D^4v'$ terms in \cref{linearised1}, all terms except the last in \cref{linearised2} and \cref{linearised3}, and all the terms in \cref{linearised4} and \cref{linearised5}. In addition, we define 
 $$ (U, T_{xx}, T_{xy}) = ( \hat U, W \hat T_{xx}, \hat T_{xy})$$
 so that all base quantities are independent of $W$. In the limit of $W \rightarrow \infty$, the reduced set of leading-order linearised equations then become
\begin{align}\label{asymptotics1}
\begin{gathered}
 \beta \hat D^4 \hat v = (1-\beta)\left[ -\hat k^2\hat D \hat \tau_{xx} +  i\hat k \hat D^2  \hat \tau_{xy} \right]
\end{gathered},
\end{align}
\begin{align}\label{asymptotics2}
\begin{gathered}
\left[i\hat k(\hat U-\hat c) + 1 \right]\hat \tau_{xx} = -\hat v \hat D \hat T_{xx} + 2i\hat k \hat T_{xx} \hat u + 2\hat T_{xy} \hat D \hat u + 2\hat \tau_{xy}\hat D\hat U
\end{gathered},
\end{align}
\begin{align}\label{asymptotics3}
\begin{gathered}
\left[i\hat k( \hat U- \hat c) + 1\right]\hat \tau_{xy} = -\hat v \hat D \hat T_{xy} + i\hat  k \hat T_{xx} \hat v + \hat  \tau_{yy}\hat D \hat U + \hat D \hat u
\end{gathered},
\end{align}
\begin{align}\label{asymptotics4}
\begin{gathered}
\left[i\hat k( \hat U- \hat c) + 1 \right] \hat \tau_{yy} = 2i\hat k \hat T_{xy} \hat v + 2 \hat D \hat v
\end{gathered},
\end{align}
\begin{align}\label{asymptotics5}
\begin{gathered}
i\hat k \hat u + \hat D \hat v = 0
\end{gathered}.
\end{align}

These equations are verified in \cref{UCM_CM} and \cref{relaminar}, where the asymptotics are valid as $W\rightarrow \infty$ across a range of $\beta$ and $\mu$. In particular, they are valid even in the UCM limit when $\beta=0$ (see \cref{UCM_CM}d). 



%
%
\section{Classification of Final States}\label{ECS}

\begin{figure} 
\centering
\begin{subfigure}[b]{0.9\textwidth} 
\includegraphics[width=\textwidth]{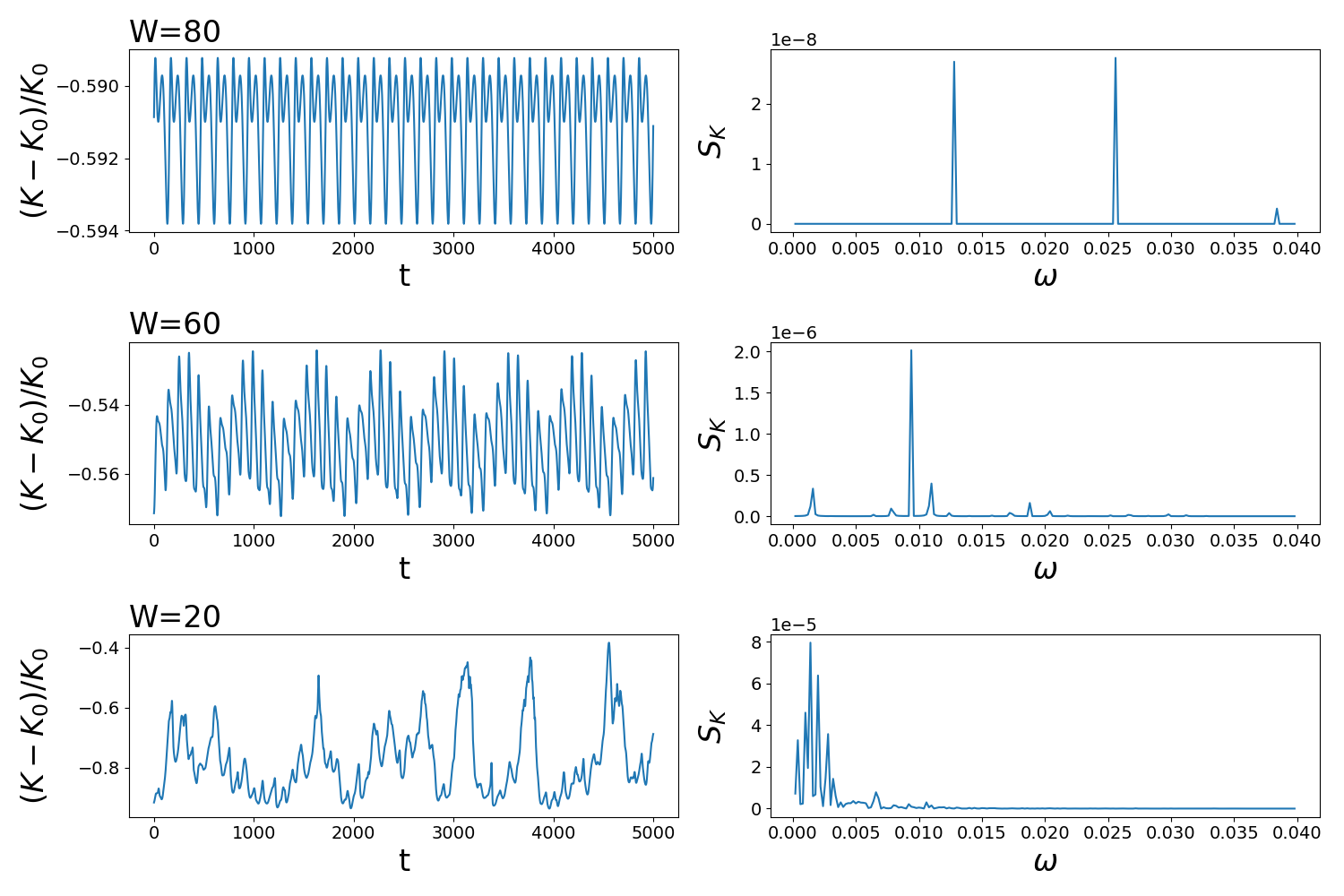}
\end{subfigure}
\caption{Time series of $K$ (left) and frequency spectra $S_K$ (right) at $W=80$ (periodic orbit), $W=60$ (quasi-periodic orbit), $W=20$ (chaos). Other parameters are $Re=0.5$, $\beta=0.95$, $\varepsilon=10^{-3}$, $n=2$ and $L_x=6\pi$. These plots correspond to three final states shown in \cref{state_diagram}b and demonstrate how the frequency spectra can be used to identified a state as a periodic orbit, quasi-periodic orbit or chaotic.}\label{frequency_spectra}
\end{figure}

To classify a final state as laminar, a travelling wave, a periodic orbit, a quasi-periodic orbit or chaotic, we consider the average kinetic energy time series, normalised with respect to the laminar base flow. Our classification protocol is as follows. If $(K-K_0)/K_0$ vanishes, then the state is laminar, while if it tends to a non-zero constant, then the state is a travelling wave. Note, this criteria does not discount equilibria, but in all cases we checked whether the identified solution was stationary on simulating to rule this out. For all fluctuating $(K-K_0)/K_0$ we looked at the frequency spectra by considering the Fourier transform,  $S_K(\omega)$ defined in (\ref{S}), of the time series.
Both periodic orbits and quasi-periodic orbits have a discrete number of frequencies with non-zero $S_K(\omega)$, with the latter having incommensurate frequencies. We characterise a state as chaotic if the frequency spectra does not consist of discrete frequencies. See \cref{frequency_spectra} for an example of periodic, quasi-periodic and chaotic solutions.

%
%

%
%
\section{Bursting Scenario with $\varepsilon=10^{-4}$}\label{transition_lower_eps}
\begin{figure} 
\centering
\begin{subfigure}[b]{0.6\textwidth} 
\includegraphics[width=\textwidth]{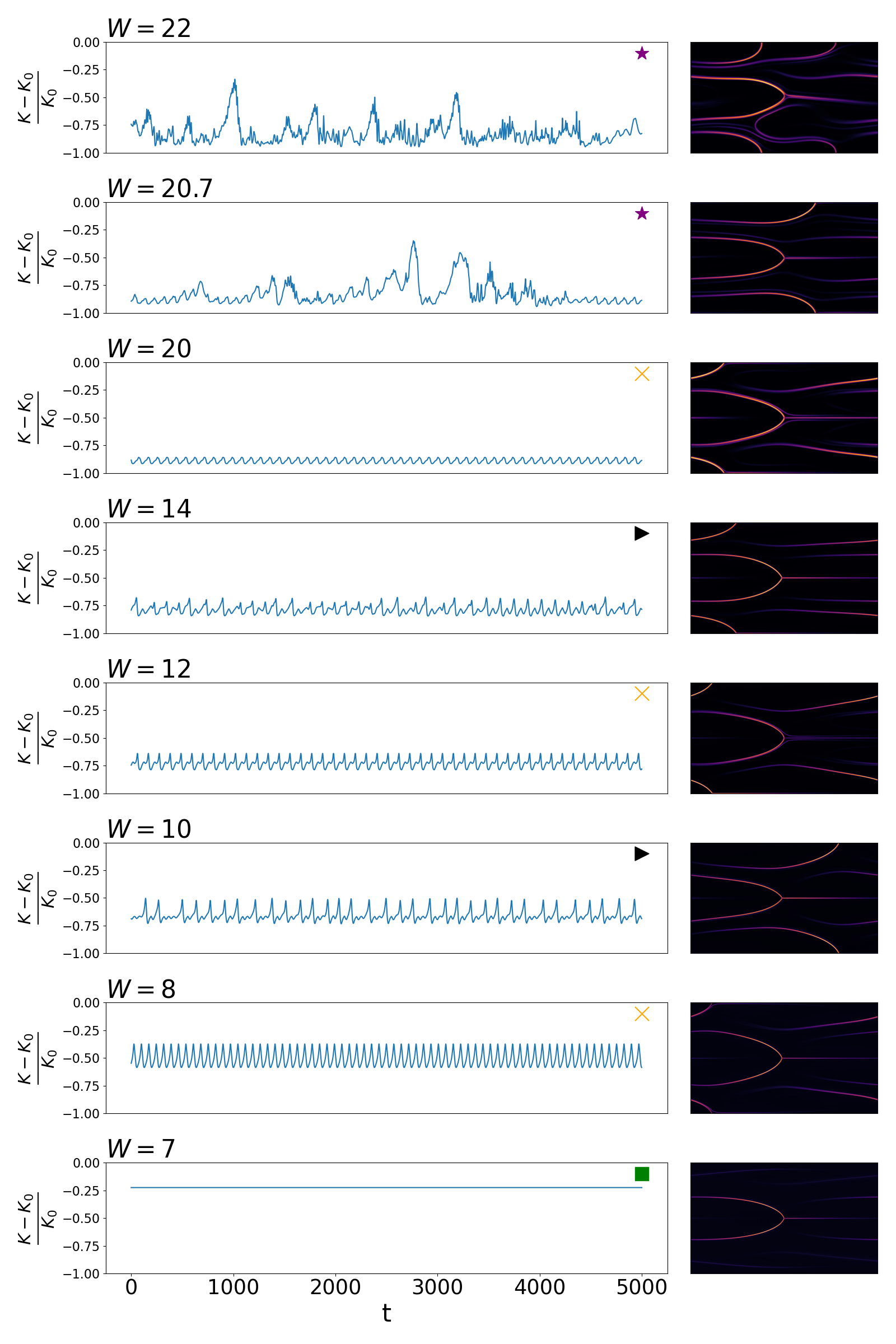}
\end{subfigure}
\caption{Kinetic energy time series of solutions (left), with the trace field at $t=5000$ (right) as $W$ is lowered from $W=22$. Parameters are $\beta=0.95$, $Re=0.5$, $\varepsilon=10^{-4}$, $n=2$ and $L_x=6\pi$. Symbols, as in \cref{state_diagram}a, show chaos (purple star), quasi-periodic orbits (black triangle), periodic orbit (yellow cross) and a travelling wave (green square). This shows that the bursting scenario connecting ET to the centre-mode arrowhead that is seen when $\varepsilon=10^{-3}$ also exists when $\varepsilon=10^{-4}$.}\label{transition2}
\end{figure}

Section \ref{centre_mode_responsible} discusses a bursting scenario leading to elastic turbulence as $W$ is varied when the polymer stress diffusion coefficient is $\varepsilon=10^{-3}$. Here, we check that the same results can be qualitatively seen when $\varepsilon=10^{-4}$. An increased resolution of $(N_x, N_y)=(384, 384)$ is used, as decreased diffusion can promote smaller scale structures, and we verify that final states are sustained when this resolution is increased to $(N_x, N_y)=(512, 512)$. The behaviour seen when $\varepsilon=10^{-3}$ (see \cref{chaos_branch_timeseries} in main text) is also seen with the reduced $\varepsilon=10^{-4}$ in \cref{transition2}, at slightly altered $W$. As a fully chaotic state at $W=22$ is tracked adiabatically down in $W$, bursting is seen ($W=20.7$), and then the state alternates between periodic orbits (e.g. $W=20$) and quasi-periodic orbits (e.g. $W=14$) before reaching a travelling wave ($W=7$). In all cases the arrowheads are thinner than when $\varepsilon=10^{-3}$.

%
%

\bibliographystyle{jfm}
\bibliography{7-bibliography}
\end{document}